\documentstyle[preprint,aps,epsf]{revtex}
\tighten
\begin{document}

\draft

\title{
Faddeev Calculations of Proton-Deuteron Radiative Capture 
with Exchange Currents 
}

\author{J. Golak, H.~Kamada$^\dagger$\footnote{present address: 
Institut f\"ur Strahlen- und Kernphysik der Universit\"at  Bonn
Nussallee 14-16, D53115 Bonn, Germany    },
         H.~Wita\l a,
        W.~Gl\"ockle$^\dagger$, J.~Kuro\'s, R.~Skibi\'nski, 
        V.~V.~Kotlyar$^\S$}
\address{Institute of Physics, Jagellonian University,
                    PL 30059 Cracow, Poland}
\address{$^\dagger$ Institut f\"ur Theoretische Physik II,
         Ruhr Universit\"at Bochum, 44780 Bochum, Germany}
\address{$^\S$ National Science Center, ``Kharkov Institute of Physics and Technology'', \\
Institute of Theoretical Physics,
Kharkov 310108, Ukraine}

\author{ K. Sagara, H. Akiyoshi$^\ddagger$
 }

\address{Department of Physics, Kyushu University, Hakozaki, Fukuoka 812, Japan}
\address{$^\ddagger$ RIKEN, 2-1 Hirosawa, Wako, Saitama 351-0198, Japan}

\date{\today}
\maketitle

\begin{abstract}
pd capture processes at various energies have been analyzed based on
solutions of 3N-Faddeev equations and using modern NN forces.
The application of the Siegert theorem is compared to the explicit use 
of $\pi$- and $\rho$-like exchange currents connected to the AV18 
NN interaction. 
Overall good agreement with cross sections and spin observables 
has been obtained
but leaving room for improvement in some cases. Feasibility studies for 3NF's 
consistently included in the 3N continuum and the 3N bound state have been 
performed as well.
\end{abstract}

\narrowtext

\section{Introduction}
\label{secIN}

Recently new nucleon-nucleon (NN) potentials have been worked out, 
which describe 
the rich NN data base below the pion threshold perfectly well. This  set of new
potentials, often called a new generation of interactions, 
comprises Nijm~I and Nijm~II~\cite{ref01},
AV18~\cite{ref02} and CD-Bonn~\cite{ref03}. 
Used in three-nucleon (3N) scattering calculations they describe many 
observables rather well, especially at lower nucleon lab energies below 
about 100 MeV~\cite{ref1}. An important insight thereby is the robustness 
of that picture: the four NN force predictions are very close 
together. This allows, in case of discrepancies to data, to think that very 
likely one will see three-nucleon force (3NF) effects. Indeed there 
is increasing evidence 
that certain discrepancies of data and NN force predictions can be cured 
by adding 3NFs~\cite{taipei-hw}. 
Good examples are the minima of the differential cross 
sections in elastic pd scattering at intermediate energies~\cite{ref2} 
or the deuteron vector analyzing power in the same process~\cite{ref3}. 
There are also counter examples which clearly demonstrate that
the correct spin structure of 3NFs has not yet been fully  established. 
A most prominent example is the analyzing power in elastic pd scattering 
at low and intermediate energies~\cite{ref4,ref5}.
In any case these results form a solid basis to study electron scattering 
and photodisintegration of $^3$He or pd-capture processes, since the dynamics 
before and after the photon has been absorbed is fairly well under control. 
Though 3NF effects are of great interest and very likely certain signatures 
have already been identified as pointed out, by far the most  
3N observables can to a large extent be well described 
by NN forces only. Also in view of applications to electromagnetic processes
it is very important to say that 
the $^3$He bound state as well as the 3N scattering states 
can reliably be calculated for one and the same Hamiltonian based on that 
set of new realistic NN forces. 

Now the new dynamical ingredient for  electromagnetic processes is the nuclear 
electromagnetic current operator. Its dependence on the electromagnetic 
nucleon form factors in case of the single nucleon current and its consistency 
to the chosen nuclear forces are of central interests. Thus the nuclear matrix 
elements depending on the electromagnetic current operator
allow the access to the electromagnetic neutron form factors in case 
of polarized $^3$He targets~\cite{ref6,ref7,ref75} and mesonic exchange 
currents (MECs)
provide important insight into the nuclear dynamics.

In this article we concentrate on effects of MECs for the 3N system, 
which for the bound and scattering states is fully treated 
by solving corresponding Faddeev equations~\cite{ref8,ref9}. 
Our techniques to handle MECs adapted to our 
general formalism is described in \cite{Kotlyar}. For low photon 
momenta and processes with real photons the Siegert theorem in a long 
wavelength approximation~\cite{ref10} is
quite popular to include some of the two-body currents. We shall also use it 
in a form which does not rely on such an approximation and we shall 
compare those results to calculations using directly MECs. 
As expected the comparison will look different for low and high energies.  

In the long wavelength approximation the Siegert theorem allows one
to write the interaction Hamiltonian in terms of the strengths of electric and magnetic
fields, i.e., in a gauge independent form. The Siegert theorem has been extended
before~\cite{Siegertext} beyond the long wavelength approximation. Diverse aspects 
and properties of the Siegert--like transformations are studied 
in the above mentioned articles and in~\cite{Siegertoth}.

We shall compare our theory to several recent and older data.
Among them are precise analyzing power data taken at Kyushu University tandem
accelerator laboratory by using an intense $d$-beam and a hydrogen gas target
of fairly high pressure and by detecting $^3$He particles instead of
$\gamma$ rays in order to simultaneously measure the whole angular
distribution~\cite{Sagara}. 

In theory this field has been actively investigated before by many groups. 
For the older calculations we refer to~\cite{ref11} for an overview. 
More recently Torre~\cite{ref12} performed for a pd capture reaction 
a very advanced calculation treating initial and final 3N states correctly
and including MEC's. Ishikawa and Sasakawa~\cite{ref13} even included 
3N forces for such a process. Further, based on finite rank approximations 
of NN forces Fonseca and Lehman~\cite{ref14} studied intensively various
pd capture reactions. At very low energies there exist highly advanced 
pd capture calculations by Friar {\em et al.}~\cite{ref15} and
Kievsky {\em et al.}~\cite{ref16}, which has been reviewed recently 
in~\cite{ref17}. More recently new investigations on photodisintegration 
and pd-capture appeared for very low energies in~\cite{Wulf,Viviani}.

Our aim is to use modern forces, keep at least approximate consistency 
between MEC's and the nuclear forces according to the Riska 
prescription~\cite{ref18}, analyze more data also at intermediate
energies and compare the results based on Siegert theorem and on
the explicit use of MEC's. We also do not restrict ourselves to low multipoles 
like in most of the previous work.

In section~\ref{secII} we rephrase the Siegert theorem working only 
in momentum space. We think this is a very transparent and algebraically 
simple notation in contrast to the rather tedious algebraic steps usually 
presented in a configuration space notation~\cite{ref19}. It also does not
require long wavelength approximations. In that context we also want to show 
the connection between the partial wave representation of our former 
work~\cite{ref21} 
and the multipole expansion. There are many kinds of MEC's. In this article 
we concentrate on $\pi$- and $\rho$-like exchange currents linked to the specific 
AV18 NN force. This is outlined in section~\ref{secIII}.
Then in section~\ref{sec3NF} we show our way to include a 3NF force into 
pd capture calculations.
The broad set of numerical results compared to data is shown 
in section~\ref{secIV}. In a future work we shall compare the different
modern NN force predictions with properly related MEC's in order to investigate
whether this dynamical scenario is robust against interchanges
of the forces. 
Finally we summarize in section~\ref{secV}.

\section{Siegert Theorem and Multipole Expansion}
\label{secII}

Multipole expansions have a long history and are a natural tool to characterize
radiative transitions between the numerous levels of nuclei. In the 3N system 
there is only one bound state for $^3$He($^3$H) and thus no obvious need for 
that notation. Also working not only at very low energies very many multipoles 
are involved and we hardly  used that notation up to now, see however~\cite{ref20}. 
The Siegert theorem~\cite{ref10} is embedded in that notation 
of multipoles and that theorem
is very powerful in the description of pd capture or the photodisintegration processes.
Therefore in order to apply it we want to exhibit the connection between our partial wave 
decomposition used up to now~\cite{ref21} to the multipole expansion. 
At the same time we want to present a short outline
leading to Siegert's theorem carried through in momentum space.
Work along this line has been presented before in \cite{elba98,blast}.

The nuclear matrix element for photodisintegration of $^3$He is
\begin{eqnarray}
{\cal J} _\xi (\vec Q) = < \overrightarrow P' \Psi_{f}^{(-)} \vert \overrightarrow \epsilon_{\xi}
(\overrightarrow Q )\cdot \overrightarrow j (0) \vert \Psi_{^3He}
  \overrightarrow P > &\equiv& \overrightarrow
\epsilon_{\xi}(\overrightarrow Q )
\cdot \overrightarrow I(\overrightarrow Q ) {\rm ,}
\label{eq:eq1}
\end{eqnarray}
where $\Psi_{^3He}$ and $\Psi_{f}^{(-)}$ are 3N bound and scattering states, 
$\overrightarrow P$ and $\overrightarrow {P'}$ the total 3N momenta 
before and after photon absorption, $\overrightarrow \epsilon_{\xi}(\overrightarrow Q )$ 
a spherical component of the photon polarization vector and
$\overrightarrow {j (0)} $ the nuclear current operator. 
It consists in general of a single nucleon part and more than one-nucleon parts. 
The three components of the nuclear matrix element 
$\overrightarrow I(\overrightarrow Q ) $  can be expanded into spherical 
harmonics. Choosing the photon direction $ \hat Q $ to point into 
the $ \hat z$-direction we expand the product of the polarization 
vector $\overrightarrow \epsilon_{\xi}(\overrightarrow z )$ and the spherical harmonics 
$Y_{l0}({\hat {Q'}})$ into vector spherical harmonics. This then leads 
immediately to

\begin{eqnarray}
\overrightarrow \epsilon_{\xi}(\overrightarrow z )
\cdot \overrightarrow I( \vert \overrightarrow Q \vert \hat z ) 
&=&
{1\over{2\sqrt{\pi}}} \sum_{J \ge1} \sum_{\l=J,J\pm 1}
\sqrt{2l  + 1}\ C(l 1 J,0 \xi) \int d\hat {Q'}
\overrightarrow Y_{J l 1}^{\xi}(\hat {Q'}) \cdot
\overrightarrow I(\vert \overrightarrow Q \vert \hat {Q'} )  \cr
&\equiv&
- \sqrt{2 \, \pi} \sum_{J \sigma} \xi^{\sigma} \sqrt{2J  + 1} \ 
T^{\sigma}_{J \xi} (\vert \overrightarrow Q \vert)
{\rm ,}
\label{eq:eq2}
\end{eqnarray}
which is the usual multipole expansion. Inserting the Clebsch-Gordan 
coefficients one gets

\begin{eqnarray}
T^{\sigma=0}_{J \xi} (Q) &\equiv& T^{el}_{J \xi} (Q) \cr
&=&
-{1\over{4\pi}}
\int d\hat {Q'} \  \lbrace
\sqrt{{J } \over { 2 J+1 }}
\overrightarrow Y_{J J+1 1}^{\xi}(\hat {Q´}) +
\sqrt{{{J+1}\over{2J+1}}}
\overrightarrow Y_{J J-1 1}^{\xi}(\hat Q') \rbrace \cdot
\overrightarrow I(\vert \overrightarrow Q \vert \hat {Q'} )
\label{eq:eq3}
\end{eqnarray}
and
\begin{eqnarray}
T^{\sigma=1}_{J \xi} (Q) \equiv T^{mag}_{J \xi} (Q)
= {1 \over {4 \pi }} \int d\hat {Q'}
\overrightarrow Y_{J J 1}^{\xi}(\hat {Q'}) \cdot
\overrightarrow I(\vert \overrightarrow Q \vert \hat {Q'} )
\label{eq:eq4}
\end{eqnarray}
This separates the magnetic multipoles 
$T^{mag}_{J \xi}$
($l=J$) and the electric
multipoles
$T^{el}_{J \xi} $
 ($l=J\pm1$). 
Now we use the identity~\cite{ref22}
\begin{eqnarray}
\hat Q Y_{J \xi }(\hat Q)  &=& -\sqrt{{{J +1}\over{2J +1}}}
\overrightarrow Y_{J J+1 1}^{\xi}(\hat Q) +
\sqrt{{{J }\over{2J+1}}}
\overrightarrow Y_{J J-1 1}^{\xi}(\hat Q),
\label{eq:eq45}
\end{eqnarray}
which allows to rewrite the term 
with $\overrightarrow Y_{J J-1 1}^{\xi}(\hat Q)$ in terms of      
$\overrightarrow Y_{J J+1 1}^{\xi}(\hat Q) $ and most importantly 
the term $\hat Q Y_{J \xi }(\hat Q)$. 
On the other hand $ \hat Q \cdot {\overrightarrow I} (\overrightarrow Q) $
occurs in the continuity equation for the electromagnetic current

\begin{eqnarray}
\overrightarrow Q \cdot \overrightarrow I( \overrightarrow Q ) &=&
 <\overrightarrow P'
\Psi_{f'}^{(-)} \vert
\lbrack H,\hat \rho(0) \rbrack \vert \Psi_{^3He}
\overrightarrow P>= <\overrightarrow P'
\Psi_{f'}^{(-)} \vert  ( E' \hat \rho(0) - \hat \rho(0) E) \vert \Psi_{^3He}
\overrightarrow P>\cr
&=& \omega
<\overrightarrow P'  \Psi_{f'}^{(-)} \vert \hat \rho(0)  \vert \Psi_{^3He}
\overrightarrow P> \equiv Q \rho(\overrightarrow Q)
\label{eq:eq5}
\end{eqnarray}
with

\begin{eqnarray}
\vert \vec Q \vert = Q=  \omega .
\end{eqnarray}
Therefore we end up for the electric multipoles with

\begin{eqnarray}
&{}&T^{el}_{J \xi} (Q) \equiv
\cr
&{}&-{1\over{4 \pi}}
 \int d\hat Q' \lbrace \sqrt{{{2J+1}\over{J}}}
\overrightarrow Y_{J J+1 1}^{\xi}(\hat Q')
\cdot \overrightarrow I (\vert \overrightarrow Q \vert \hat Q')
 +\sqrt{{{J+1}\over{J}}}  Y_{J\xi}(\hat Q')
\rho (\vert \overrightarrow Q \vert \hat Q') \rbrace.
\label{eq:eq6}
\end{eqnarray}
Another derivation of that formula is presented in the appendix.
Note that the first term in the curly bracket in Eq.~(\ref{eq:eq6}) 
is normally neglected in a long wavelength
approximation. The last term is now the density 
matrix element, which is believed to be less affected (at least for 
low momenta) by two-body effects. Therefore a single body operator might be 
a reasonable approximation. In this article we use that assumption 
and keep also the first term, however in a single nucleon current 
approximation. In the result section we refer by "Siegert" to such a choice.
Thus the identity given in Eq.~(\ref{eq:eq45})
together with current conservation and exact 3N eigenstates 
of the Hamiltonian $H$ allowed to
shift effects of the unknown current matrix element 
$\overrightarrow I(\overrightarrow Q) $ into       
$\rho(\overrightarrow Q ) $ and into higher multipoles.
Clearly this is only the first step towards introducing two-body currents. 

Now let us indicate in the example of the single nucleon density operator 
the connection between our partial wave expansion and the multipole expansion. 
Working in momentum space and in a partial wave decomposition for 3 nucleons 
we use the basis $ \vert p q \beta >$, where $p$ and $q$ are
the magnitudes of Jacobi momenta and $ \beta $ a set of discrete quantum 
numbers (orbital and spin angular momenta coupled to the total 3N angular 
momentum and isospin quantum numbers)~\cite{ref21}.

The density matrix element has the structure 
\begin{equation}
\rho ( \vec Q )
\ = \ {\sum \hskip -13 pt \int} 
< \Psi_{f}^{(-)} \vert p q \beta > \, < p q \beta \vert \hat \rho ( \vec Q ) \vert \Psi_{^3He} >
\end{equation}
For our purpose it is sufficient to regard only
$ < p q \beta \vert \hat \rho ( \vec Q ) \vert \Psi_{^3He} > $,
which is shown in~\cite{ref21} to have the form for a single nucleon 
density operator

\begin{eqnarray}
<p  q  \beta   ({\cal J} M) \vert \hat \rho (\vec Q \parallel \hat z ) \vert \Psi_{bound} ({1 \over 2} M')  >
 =  \delta _{M, M'} \delta_{M_T,M_{T'}} { 1\over 2}  \sqrt{ \hat {\cal J}} \sqrt{\hat L } (-1)^ {S+ {\cal J} } \cr 
[ I ^{(p)} (t  ,T, M_T ) F_1^p (\vec Q)
+  I ^{(n)} (t  ,T, M_T ) F_1^n (\vec Q)]
\sum_{\beta '} \delta_{l,l'} \delta_{s,s'} \delta_{S,S'} \delta_{t,t'} \sqrt{\hat {\lambda '} \hat{L'}}
\cr \times
{}\sum _{ \lambda_1 + \lambda_2 = {\lambda '}} \sqrt{ { (2 {\lambda '} + 1) ! } \over { (2\lambda_1) ! (2\lambda_2) !} }
\ q^{\lambda_1} ( {2 \over 3} Q ) ^{\lambda _2}
\cr \times
{}\sum_k \hat k \ C(\lambda_1 , k , \lambda ; 0,0,0)
\int _{-1} ^1 dx P_k (x) { < p \tilde q \beta ' \vert \Psi_{bound} > \over {\tilde q}^{\lambda '} }
\cr \times
{} \sum _g \sqrt{ \hat g }\ C( \lambda _2 , k , g ; 0 ,0,0)
\left\{  \matrix{ \lambda_2 & \lambda _1 & \lambda ' \cr
                   \lambda  &  g         & k        \cr       }  \right\}
\left\{  \matrix{  l  & \lambda  & L  \cr
                   g  &  L '       & \lambda '      \cr       }  \right\}
\cr \times
{}
\left\{  \matrix{ {\cal J}  & L   & S  \cr
                  L ' &  {1 \over 2}         & g        \cr       }  \right\}
\ C({\cal J} , g , { 1 \over 2} ; M , 0 , M ').
\label{eq:eq7}
\end{eqnarray}
We also indicated now in the bra vector explicitely the total 3N angular
momentum ${\cal J}$ and its magnetic quantum number $M$.
Here it is assumed that $ \hat Q = \hat z $. On the other hand the part 
of the electric multipole related to $\rho $ in Eq.~(\ref{eq:eq6}) is
\begin{eqnarray}
T^{el;\rho}_{J \xi} &\equiv&  {1\over{4 \pi}}
 \int d\hat Q'
\sqrt{{{J+1}\over{J}}}  Y_{J \xi}(\hat Q')
\rho(\vert \overrightarrow Q \vert \hat Q').
\label{eq:eq8}
\end{eqnarray}

In order to perform the integral in Eq.~(\ref{eq:eq8}) we need 
the generalization of Eq.~(\ref{eq:eq7}) to an arbitrary $\hat Q$-direction. 
This is easily achieved by replacing in Eq.~(\ref{eq:eq7})
$ \delta_{M, M'} \ C({\cal J} , g , { 1 \over 2} ; M , 0 , M ') $ 
by
\begin{eqnarray}
\sqrt{ 4 \pi \over \hat g} \, Y_{g, M'-M} \, ( \hat Q ) C({\cal J} , g , { 1 \over 2} ; M , M'-M , M') 
\label{eq:eq9}
\end{eqnarray}
Then the integral can be performed and one ends up with

\begin{eqnarray}
T^{el;\rho}_{J \xi}
\ = \ {\sum \hskip -13 pt \int} 
< \Psi_{f}^{(-)} \vert p q \beta > \, < p q \beta \vert \Phi^{el;\rho}_{J \xi} >
\label{eq:eq9.5}
\end{eqnarray}
where $< p q \beta \vert \Phi^{el;\rho}_{J \xi} > $ is given by 

\begin{eqnarray}
< p q \beta \vert \Phi^{el;\rho}_{J \xi} > = \cr
\delta_{M_T,M_{T'}} \, { 1\over 4} \, { 1 \over {\sqrt \pi}} \, \sqrt{{{J+1}\over{J}}} \, (-1)^{\xi} \, 
C({\cal J} , J , { 1 \over 2} ; M , -\xi , M') \cr
\left\{  \matrix{  l  & \lambda  & L  \cr
                   J  &  L'         & \lambda '       \cr       }  \right\} \,
\left\{  \matrix{ {\cal J}  & L  & S  \cr
                  L '  &  {1 \over 2}    & J        \cr       }  \right\} \,
\sqrt{ \hat {\cal J}} \sqrt{\hat L } (-1)^ {S+ {\cal J} } \cr
[ I ^{(p)} (t  ,T, M_T ) F_1^p (\vec Q) + I ^{(n)} (t ,T, M_T ) F_1^n (\vec Q)] \cr
\sum_{\beta '} \ \delta_{l,l'} \delta_{s,s'} \delta_{S,S'} \delta_{t,t'} \sqrt{\hat {\lambda '} \hat {L '} } \cr
\sum_{\lambda_1 + \lambda_2 = {\lambda '} } \sqrt{ { (2 {\lambda '} + 1) ! } \over { (2\lambda_1) ! (2\lambda_2) !} }
\ q^{\lambda_1} ( {2 \over 3} Q ) ^{\lambda _2} \cr
\sum_k \, \hat k \ C(\lambda_1 , k , \lambda  ; 0,0,0) \, C( \lambda_2 , k , J ; 0 ,0,0) \,
\left\{  \matrix{ \lambda_2 & \lambda _1 & \lambda ' \cr
                   \lambda  &  J         & k        \cr       }  \right\} \cr
\int _{-1} ^1 dx P_k (x) { < p \tilde q \beta ' \vert \Psi_{bound} > \over {\tilde q}^{\lambda '} } .
\label{eq:eq10}
\end{eqnarray}
As a consequence the variable $g$ in Eq.~(\ref{eq:eq7}) has to be identified 
with $J$ -- the multipole order of Eq.~(\ref{eq:eq10}). A corresponding relation can easily 
be worked out for the remaining parts in Eq.~(\ref{eq:eq6}) and Eq.~(\ref{eq:eq4}).

\section{Model of Meson Exchange Currents}
\label{secIII}

\begin{figure}[h!]
\epsfbox{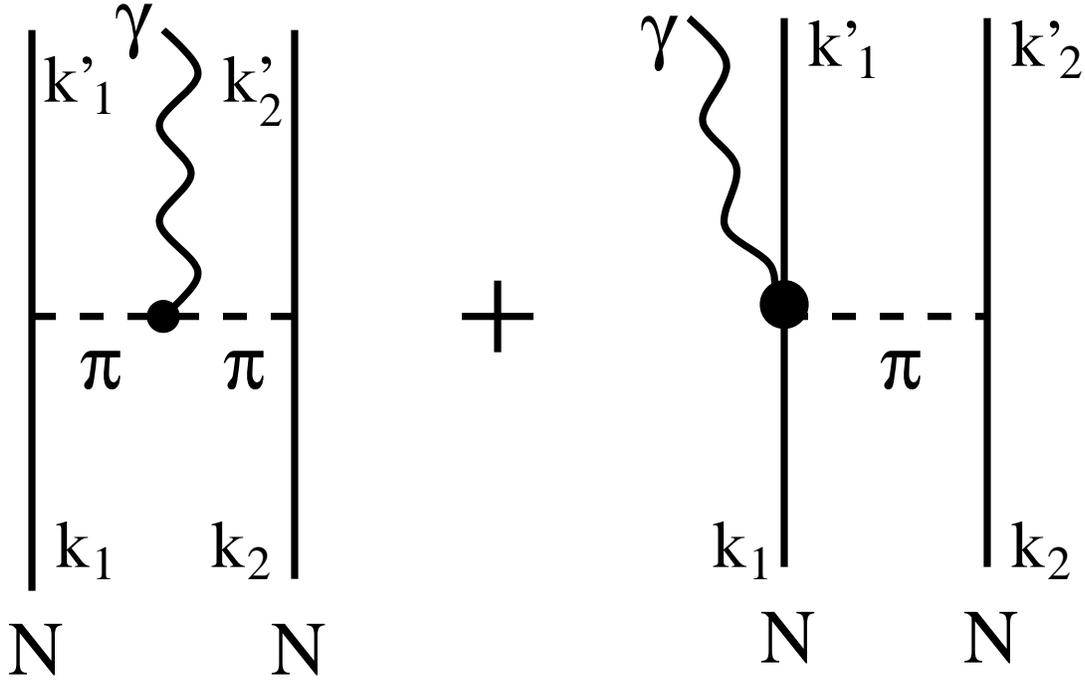}
\caption[ ]{The "pion-in-flight" and "seagull" pionic exchange currents.}
\label{fig1}
\end{figure}

The study of meson-exchange currents has also a long history. We follow 
the scheme adopted by the Urbana-Argonne collaboration~\cite{ref17}, based on the Riska 
prescription~\cite{ref18}. Dominant contributions arise 
from the $\pi$- and $\rho$-exchanges, 
to which we restrict ourselves in this article. The $\pi$-current
consists of the so called seagull and pion in flight parts, which 
in case of a true pion exchange are displayed in Fig.~\ref{fig1}. The well 
known expressions are

\begin{eqnarray}
\vec j^{seagull}(\vec p_1 , \vec p_2 ) 
&\equiv& 
 i [\tau (1) \times \tau (2) ] _z
F_1 ^V \lbrack v_\pi (p_1)( \vec \sigma (1) \cdot 
\vec p_1 ) \vec \sigma (2) - v_\pi (p_2) ( \vec \sigma (2) \cdot 
\vec p_2 ) \vec \sigma (1) \rbrack ,
\cr
\vec j^{pionic} (\vec p_1 ,\vec p_2)
&\equiv& 
 i [\tau (1) \times \tau (2) ] _z 
F_1^V (\vec p_1 - \vec p_2) \lbrack \vec \sigma (1) \cdot
\vec p_1 \rbrack  \lbrack\vec \sigma (2) \cdot \vec p_2 \rbrack 
{ { v_\pi (p_2 ) - v_\pi (p_1 )  } \over { p_1 ^2 - p^2 _2 } }
\end{eqnarray}
where $\vec p_1$ and $\vec p_2$ are defined  in terms of the initial and final momenta as

\begin{eqnarray}
\vec p_1 &\equiv& \vec{ k_1} '  - \vec k_1 
\cr 
\vec p_2 &\equiv& \vec {k_2} '  - \vec k_2 
\end{eqnarray}

\noindent
Further $F_1^V$  is the isovector electromagnetic nucleon form factor. 
In case of just a pion exchange and dropping the strong form factors
$v_\pi (p) $ is given as
\begin{eqnarray}
v_\pi (p) ={{ f_{\pi NN} ^2 } \over { m_\pi ^2}} {1 \over {m_\pi ^2 
+ p^2 }}, 
\end{eqnarray}  
where $f_{\pi NN}$ and $m_\pi$ are the pseudovector $\pi NN$ coupling constant
and the pion mass, respectively.
Similarly the $\rho$-current is given as

\begin{eqnarray} 
\vec j ^{\rho }_{12} &\equiv&  \vec j_{\rho ,I} + \vec j_{\rho ,II} 
+ \vec j _{\rho ,III} + \vec j _{\rho ,IV}
 \cr
{\vec j}_{\rho, I} ( {\vec p}_1 , {\vec p}_2 ) \ &\equiv& \
  i [ {\vec \tau}(1) \times {\vec \tau}(2) ]_z \ F_1^V ( Q^2 ) \
\frac{ {\vec p}_1 - {\vec p}_2 }{ p_1^2 - p_2^2 } \
\left( v_\rho^S (p_2) - v_\rho^S (p_1) \right)
\cr 
{\vec j}_{\rho, II} ( {\vec p}_1 , {\vec p}_2 ) \ &\equiv& \
 - \  i [ {\vec \tau}(1) \times {\vec \tau}(2) ]_z \ F_1^V ( Q^2 ) \cr
 &&
\left( v_\rho (p_2) \, {\vec \sigma}(1) \times ( {\vec \sigma}(2) \times {\vec p}_2 )
 \ - \ v_\rho (p_1) \, {\vec \sigma}(2) \times ( {\vec \sigma}(1) \times {\vec p}_1 )  \right)
\cr
{\vec j}_{\rho, III} ( {\vec p}_1 , {\vec p}_2 ) \ &\equiv& \
 - \  i [ {\vec \tau}(1) \times {\vec \tau}(2) ]_z \ F_1^V ( Q^2 ) \
\cr
&&
\frac{ v_\rho (p_2) - v_\rho (p_1) }{ p_1^2 - p_2^2 } \
\left( ( {\vec \sigma}(1) \times {\vec p}_1 ) \cdot ( {\vec \sigma}(2) \times {\vec p}_2 ) \right)
 ( {\vec p}_1 - {\vec p}_2 )
\cr 
{\vec j}_{\rho, IV} ( {\vec p}_1 , {\vec p}_2 ) \ &\equiv& \
 - \  i [ {\vec \tau}(1) \times {\vec \tau}(2) ]_z \ F_1^V ( Q^2 ) \
\frac{ v_\rho (p_2) - v_\rho (p_1) }{ p_1^2 - p_2^2 } \ 
\cr
&&
\left( {\vec \sigma}(2) \cdot ( {\vec p}_1 \times {\vec p}_2 ) ( {\vec \sigma}(1) \times {\vec p}_1 )\ + \
 {\vec \sigma}(1) \cdot ( {\vec p}_1 \times {\vec p}_2 ) ( {\vec \sigma}(2) \times {\vec p}_2 ) \right) {\rm .}
\end{eqnarray}
Again for just a $\rho$-exchange one would have

\begin{eqnarray}
&&v_\rho (k) = - \left( g_{\rho NN} \over 2 m_N \right)^ 2 { (1 + \kappa )^2 \over m_\rho ^2 + k^2 }, 
\cr
&&v_\rho ^S (k) = g_{\rho NN}^2 { 1 \over m_\rho^2 + k^2 } ,
\end{eqnarray}
where $g_{\rho NN}$, $\kappa$ and $m_\rho$ are
the vector, tensor $\rho NN$ coupling constants and the $\rho$ meson mass,
respectively. 
Now according to Riska's prescription one identifies  $\pi$- and $\rho$-like parts in the given
NN potential and using the continuity equation finds a linkage of the potential to the
exchange currents. The procedure is as follows

\begin{eqnarray}
v_\pi (k) \longrightarrow v_{PS} (k) = 2 v ^ {t\tau} (k) -v ^{\sigma \tau} (k),
\end{eqnarray}
\begin{eqnarray}
v_\rho (k) \longrightarrow v_{V} (k) =  v ^ {t \tau} (k) + v ^{\sigma \tau} (k) 
\end{eqnarray}
and
\begin{eqnarray}
v_\rho^S (k) \longrightarrow v_{V}^S (k) = v^{\tau} (k),
\end{eqnarray}
where the functions $v_{PS} (k)$, $v_{V} (k)$, $v_{V}^S (k)$ are evaluated as
\begin{eqnarray}
v^{\sigma \tau} (k) = { 4 \pi \over k ^ 2} \int _0 ^\infty dr r^2 [ j_0 (kr) -1] v ^{\sigma \tau} (r),
\end{eqnarray}
\begin{eqnarray}
v^{t \tau } (k) = { 4 \pi \over k ^ 2} \int _0 ^\infty dr r^2 j_2 (kr)  v ^{ t \tau} (r),
\end{eqnarray}
\begin{eqnarray}
v^{\tau} (k) = 4 \pi \int _0 ^\infty dr r^2 j_0 (kr) v ^{\tau} (r).
\end{eqnarray}

\begin{figure}[h!]
\epsfbox{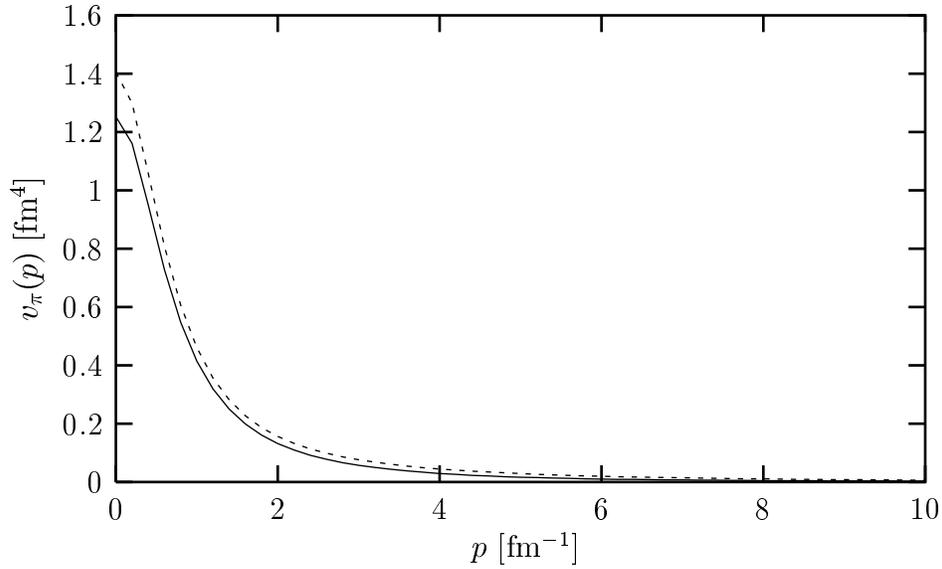}
\caption[ ]{Comparison of the form factors $v_\pi$ for AV18 (solid line) 
            and AV14 (dashed line).}
\label{fig2}
\end{figure}

\begin{figure}[h!]
\epsfbox{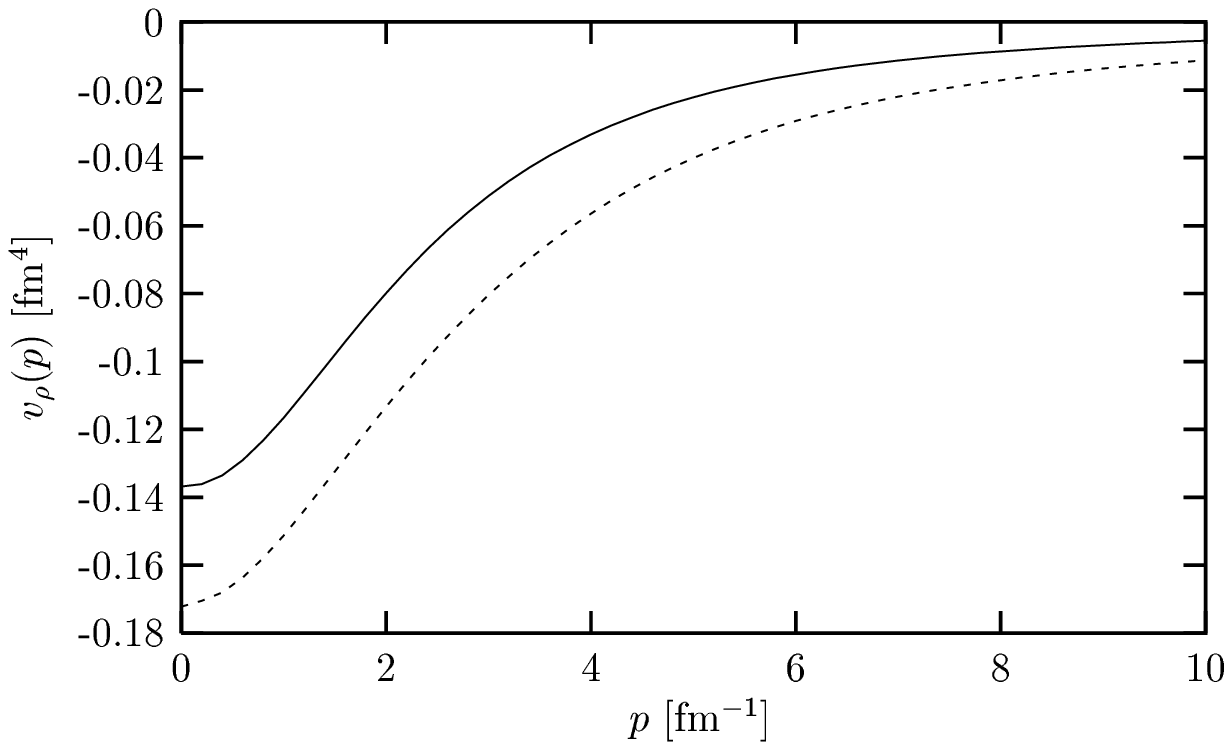}
\caption[ ]{Same as in Fig.~\protect\ref{fig2} for $v_\rho$.}
\label{fig3}
\end{figure}

\begin{figure}[h!]
\epsfbox{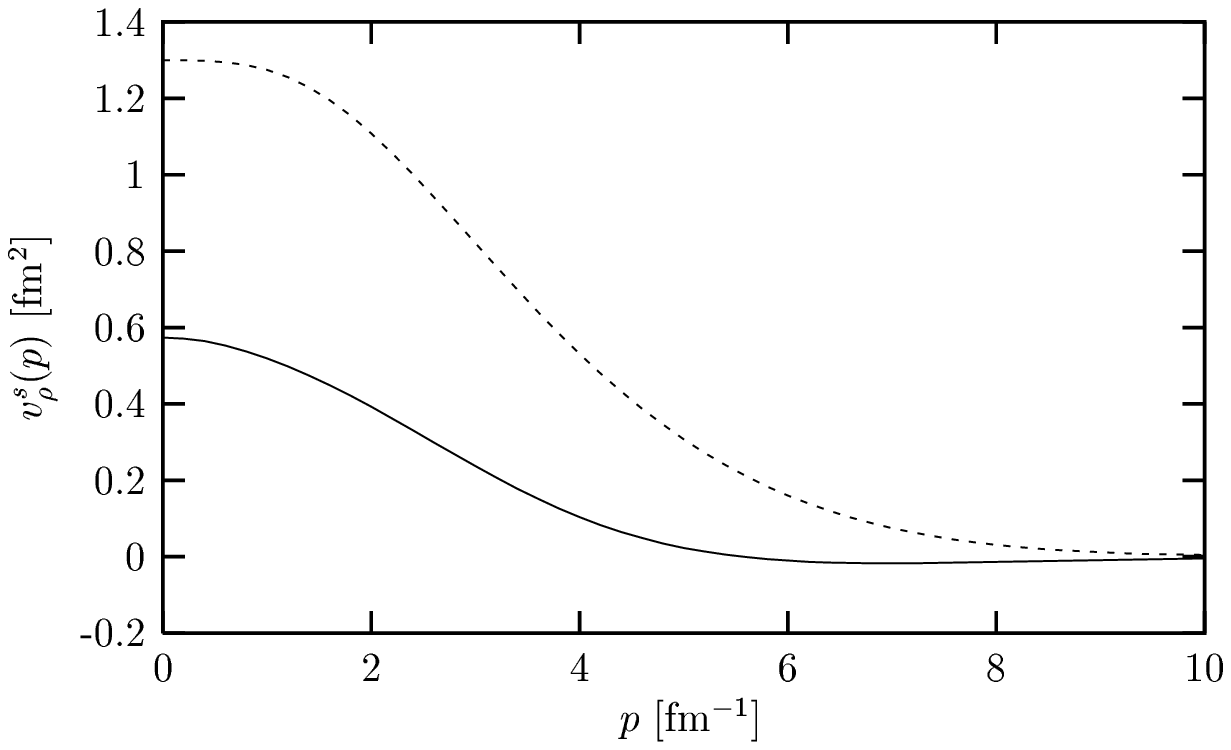}
\caption[ ]{Same as in Fig.~\protect\ref{fig2} for $v_\rho^s$.}
\label{fig4}
\end{figure}

The functions $v ^{\sigma \tau} (r)$, $v ^{ t \tau} (r)$, $v ^{\tau} (r)$ 
are taken from the AV18 potential and are the radial functions accompanying
the spin-isospin, tensor-isospin
and isospin operators, respectively.
It is interesting to look at the potential dependent 
functions $v_{PS} (k)$, $v_{V} (k)$
and $v_{V}^S (k)$ for different NN potentials. 
In Figs.~\ref{fig2}-\ref{fig4} we compare them for the AV14~\cite{refAV14} 
and AV18~\cite{ref02}.

\section{Inclusion of a Three-Nucleon Force}
\label{sec3NF}

In this section we would like to demonstrate our way of including
a three-nucleon force in pd capture calculations. For the 3N bound state 
it has been done in~\cite{ref8} and also fully converged calculations 
are available for elastic and inelastic nucleon-deuteron scattering~\cite{ref1,ref2,ref3}.

We apply the method proposed in~\cite{Newmethod} and start directly
with the nuclear matrix element for the pd capture process:

\begin{equation}
{\bar{\cal J}}_\xi (\vec Q) \equiv 
< \Psi_{^3He} \overrightarrow P \vert
\overrightarrow \epsilon_{\xi} (\overrightarrow Q )\cdot \overrightarrow j (0) 
\vert \overrightarrow P' \Psi_{f}^{(+)} >
\label{eq:3nf1}
\end{equation}

The scattering state $\Psi_{f}^{(+)}$ is then split into three 
Faddeev components, which for a system of identical particles 
reads 

\begin{equation}
\Psi_{f}^{(+)} = \psi_1 + \psi_2 + \psi_3 = (1 + P_{12} P_{23} + P_{13} P_{23}) \psi_1 \equiv (1 + P) \psi_1
\label{eq:3nf2}
\end{equation}

The Faddeev component $\psi$ fullfils the following equation (we drop the index $1$)

\begin{equation}
\psi = \phi + G_0 t P \psi + (1 + G_0 t) G_0 V_4^{(1)} (1 + P) \psi .
\label{eq:3nf3}
\end{equation}
Here $G_0$ is the free 3N propagator, $t$ the off-shell
NN t-operator and $\phi$
is a product of the deuteron state
and a momentum eigenstate of the spectator nucleon.
$V_4^{(1)}$ results from the decomposition of a 3NF into three parts
$V_4^{(i)}$, which individually are symmetrical under the exchange 
of particles $j$ and $k$ ($j \ne i$ and $ k\ne i$).
\begin{equation}
V_4 = V_4^{(1)} + V_4^{(2)} + V_4^{(3)}
\label{eq:3nf4}
\end{equation}

Introducing the amplitude $\tilde{T}$, which obeys the equation~\cite{Newmethod}
\begin{equation}
\tilde{T} = t P \phi + (1 + t G_0) V_4^{(1)} (1 + P) \phi + t P G_0 \tilde{T} 
          + (1 + t G_0) V_4^{(1)} (1 + P) G_0 \tilde{T}, 
\label{eq:3nf5}
\end{equation}
we end up with 
\begin{equation}
\psi = \phi + G_0 \tilde{T}. 
\label{eq:3nf6}
\end{equation}
Eq.~(\ref{eq:3nf5}) reduces to the form:
\begin{equation}
T = t P \phi + t P G_0 T, 
\label{eq:3nf7}
\end{equation}
when the Hamiltonian contains only NN interactions ($V_4 = 0$).

In such a way we study 3NF effects by choosing just either the solution 
of Eq.~(\ref{eq:3nf5}) or Eq.~(\ref{eq:3nf7}), inserting it into 
Eq.~(\ref{eq:3nf6}) and Eq.~(\ref{eq:3nf2})
 and applying the nuclear 
current operator to the proper 3N bound state wave function.

Thus the nuclear matrix element for the pd capture process
${\bar{\cal J}}_\xi (\vec Q)$ can be split into the plane wave part
and the part containing all initial state interactions
\[
{\bar{\cal J}}_\xi (\vec Q) \ = \
< \Psi_{^3He} \overrightarrow P \vert
\overrightarrow \epsilon_{\xi} (\overrightarrow Q )\cdot \overrightarrow j (0) 
\vert (1 + P) \phi \overrightarrow P' >
\]
\begin{equation}
\ + \
< \Psi_{^3He} \overrightarrow P \vert
\overrightarrow \epsilon_{\xi} (\overrightarrow Q )\cdot \overrightarrow j (0) 
\vert (1 + P) G_0 \tilde{T} \overrightarrow P' >
\label{eq:3nf8}
\end{equation}

\section{Results}
\label{secIV}

\begin{figure}[h!]
\epsfbox{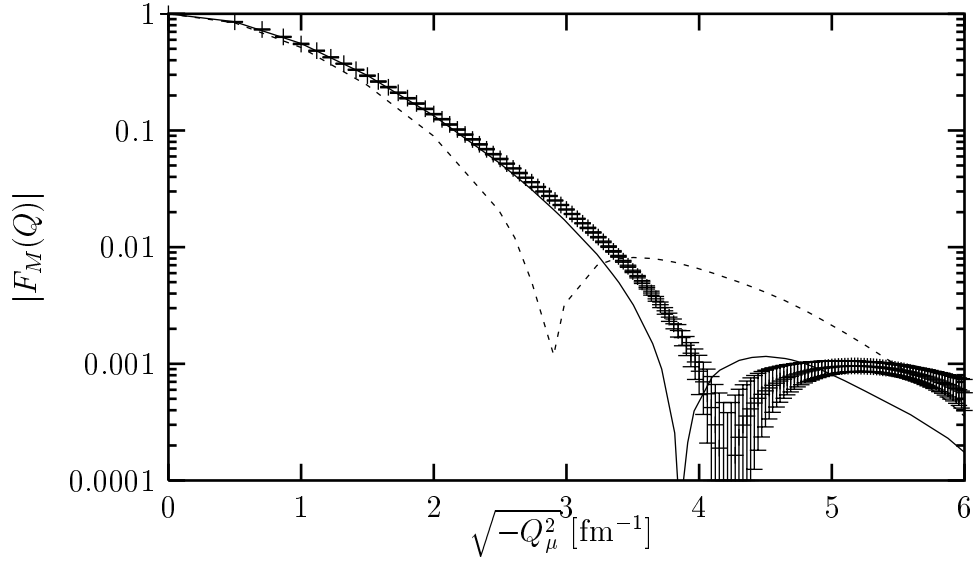}
\caption[ ]{The magnetic form factor of ${}^3$He. The solid line includes 
            MEC (see text), while the dashed line is based on a single-nucleon 
            current study. Data are from~\cite{m3hedata}.}
\label{fig5}
\end{figure}

\begin{figure}[h!]
\epsfbox{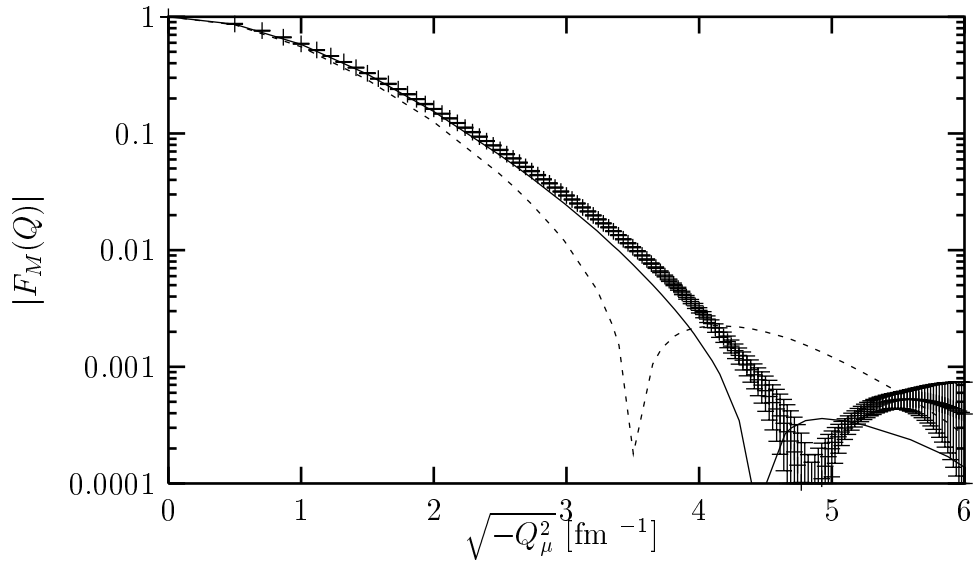}
\caption[ ]{Same as in Fig.~\protect\ref{fig5} for ${}^3$H. 
            Data are from~\cite{m3hedata}.}
\label{fig6}
\end{figure}

Let us first repeat calculations performed before by several groups
for the magnetic 
form factors of $^3$He and $^3$H. In Figs.~\ref{fig5}-\ref{fig6} we compare 
the single nucleon current prediction with the one including the $\pi$-
and $\rho$-like exchange currents. We use the AV18 potential~\cite{ref02}
 without the
various electromagnetic parts. Therefore the result of a Faddeev calculation
for the $^3$H binding energy is only 7.623 MeV. The electromagnetic nucleon 
form factors ($G_E$ and $G_M$) are from~\cite{dipole}. 
We see the well known strong shift of theory towards the data when 
including the MEC's. 
In this article we restrict ourselves to relatively low photon momenta
(see Table~\protect\ref{EpEdEgamma}), therefore we are not concerned
about the discrepancies at the higher $Q$-values 
in Figs.~\ref{fig5}--\ref{fig6}, where also relativity should play some role.
In relation to the higher $Q$-values we refer to \cite{ref17,ref175}, where theory
has been pushed further including additional MEC's and $\Delta$-admixtures.

Now we come to the main results 
and regard first the photon angular distributions for pd capture reactions. 
For the current matrix element ${\cal J}_\xi (\vec Q) $ defined
in Eq.~(\ref{eq:eq1}) the unpolarized  cross section in the total momentum zero frame
(cm) is given as
\begin{equation}
d \sigma ^ {cm} \ = \
(2\pi)^2 \alpha  { 1 \over { \vert \vec{v}_d - \vec{v}_p \vert }    } 
{ 1 \over \omega }  { 1 \over 6} \sum_{m_d,m_p,M}\sum_{\xi=\pm 1} 
\vert {\cal J} _\xi (\vec Q) \vert ^2 Q^2 dQ d\Omega_Q 
\, \delta \left( \sqrt{ m_{^3He}^2 + Q^2} + Q - \sqrt{s} \right)
\end{equation} 
where $\alpha$, $\vec{v}_d$, $\vec{v}_p$ and $\omega$ are the fine structure constant 
($\approx 1/137$), 
the velocities of the incident
deuteron and proton, and the energy of the outgoing photon, respectively.
$\sqrt{s}$ is the total energy of the p+d and $^3$He+$\gamma$ systems
($ \sqrt{s} = E_d + E_p = E_{^3He} + Q $).
The differential cross section ($d \sigma / d \Omega_Q$) is then obtained as
\begin{eqnarray}
\left( { d \sigma \over d \Omega_Q } \right)^{cm} = (2\pi)^2 \alpha 
{ 1 \over 6} \sum_{m_d,m_p,M}\sum_{\xi=\pm 1} \vert {\cal J} _\xi (\vec Q) \vert ^2 
\  { { E_p \, E_d  \, E_{^3He} \, Q } \over { s \, p_p } } ,
\end{eqnarray}
where $p_p$ is the magnitude of the proton momentum.
Note that we use the relativistic phase space factor.

\begin{figure}[hbt]
\epsfbox{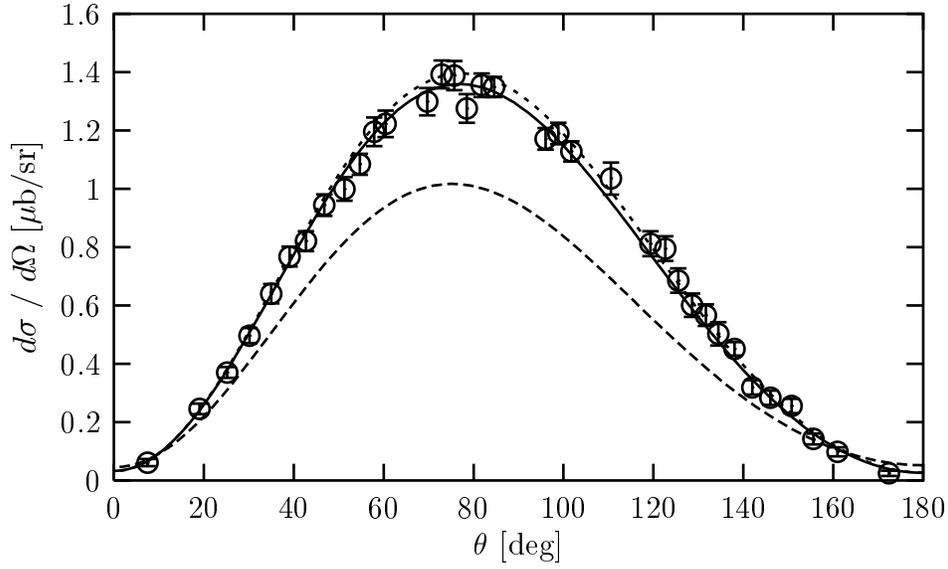}
\caption[]{The photon angular distribution for pd capture at $E_d$= 19.8 MeV
           against the c.m. $\gamma$-p scattering angle.
           The single nucleon current prediction is given by the dashed line,
           including MEC leads to the solid line and the dotted line
           is due to Siegert. Data are from~\protect\cite{Belt}.}
\label{fig75}
\end{figure}

\begin{figure}[hbt]
\epsfbox{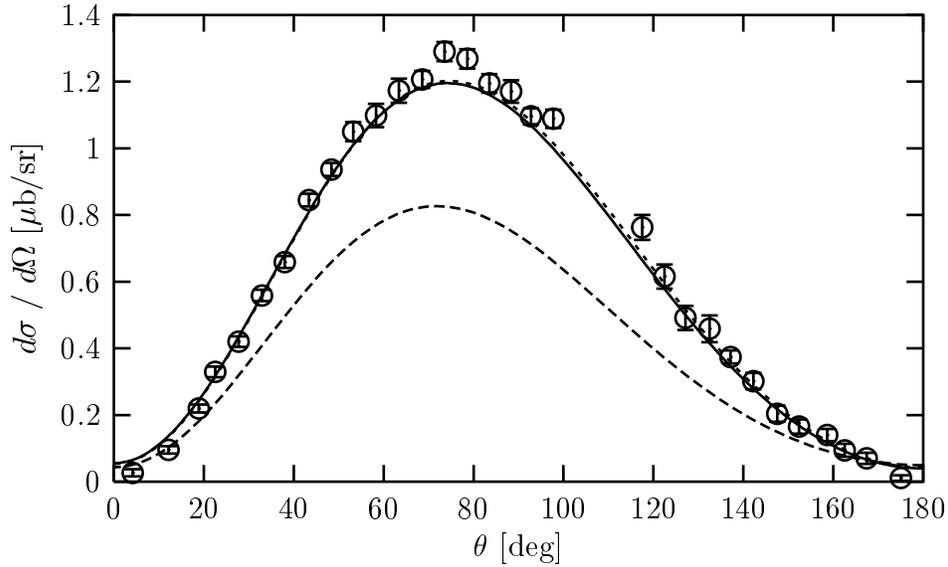}
\caption[]{The same as in Fig.~\protect\ref{fig75} at $E_d$= 29.6 MeV.
           Note that the calculations are for $E_d$= 29.2 MeV.
           Data are from~\protect\cite{Belt}.}
\label{fig76}
\end{figure}

\begin{figure}[hbt]
\epsfbox{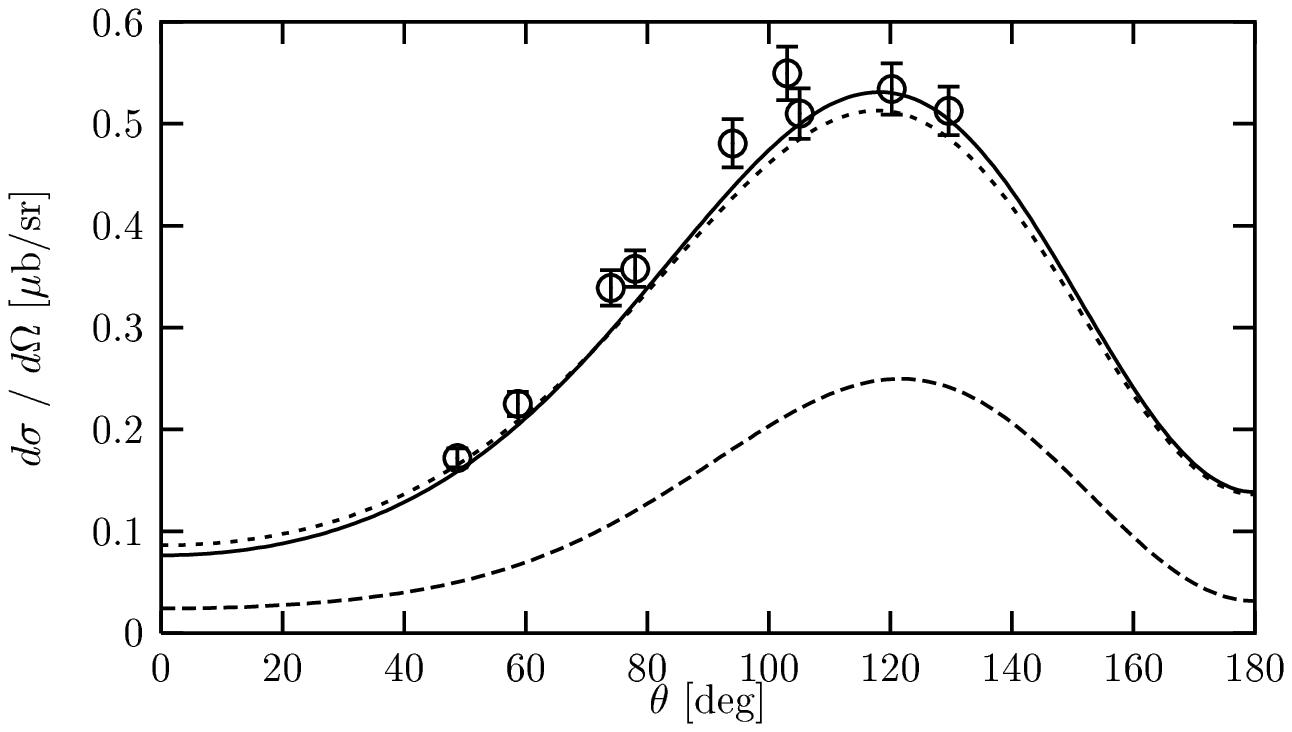}
\caption[]{The same as in Fig.~\protect\ref{fig75} at $E_d$= 95 MeV.
           $\theta$ is the c.m. $\gamma$-d scattering angle.
           Data are from~\protect\cite{Pitts}.}
\label{fig8}
\end{figure}

\begin{figure}[hbt]
\epsfbox{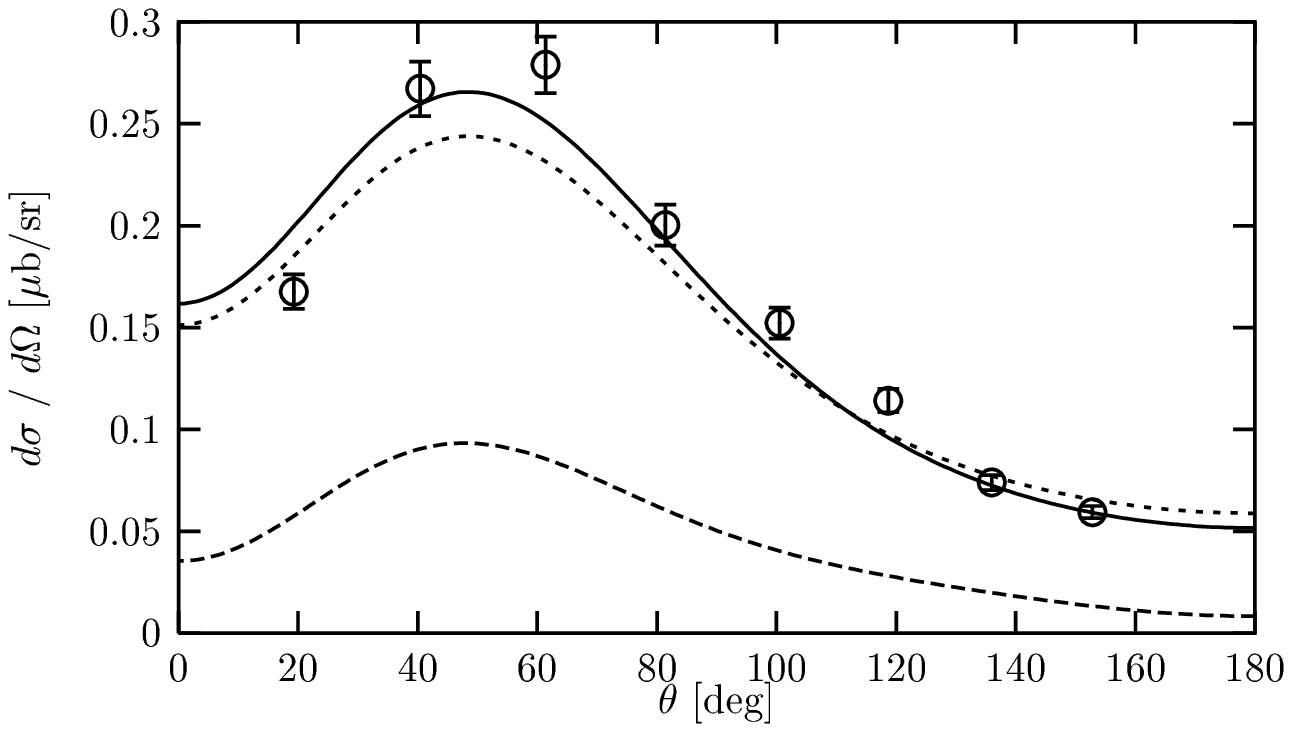}
\caption[]{The same as in Fig.~\protect\ref{fig75} at $E_p$= 100 MeV.
           $\theta$ is the c.m. $\gamma$-p scattering angle.
           Data are from~\protect\cite{Pickar}.}
\label{fig9}
\end{figure}

\begin{figure}[hbt]
\epsfbox{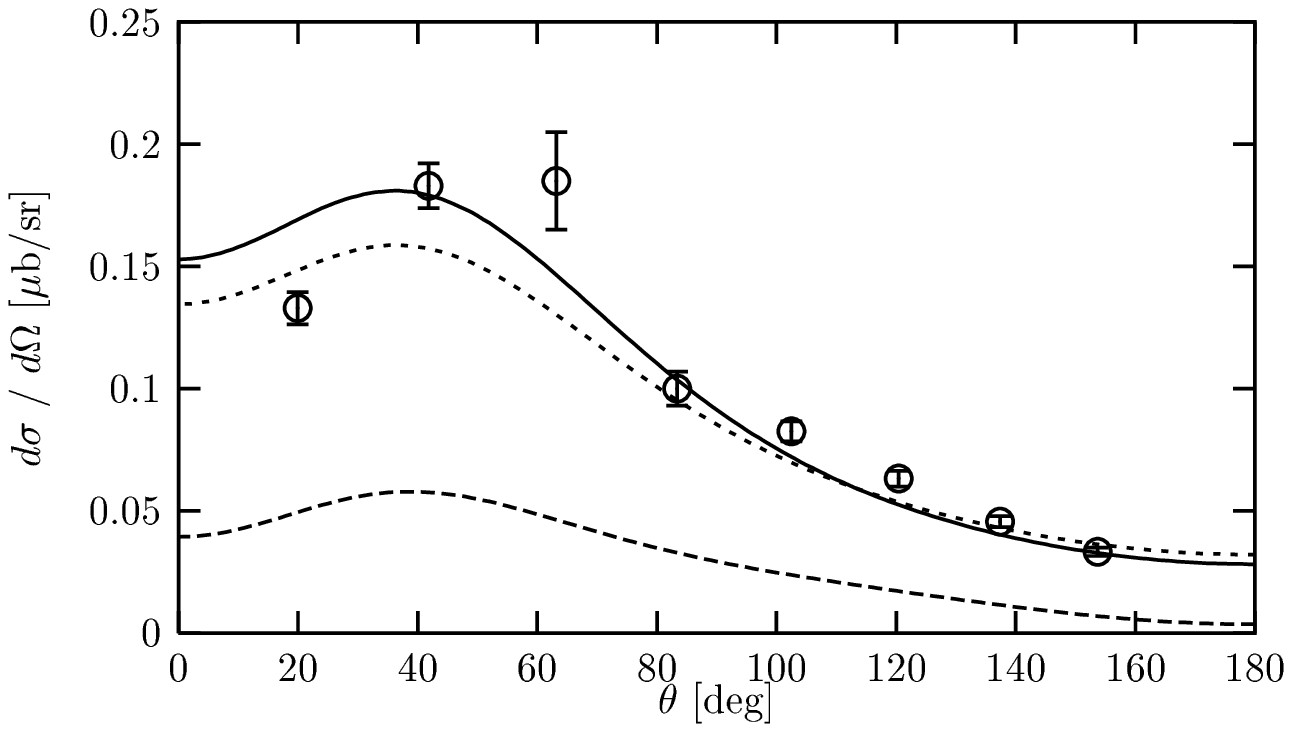}
\caption[]{The same as in Fig.~\protect\ref{fig75} at $E_p$= 150 MeV.
           $\theta$ is the c.m. $\gamma$-p scattering angle.
           Data are from~\protect\cite{Pickar}.}
\label{fig10}
\end{figure}

\begin{figure}[hbt]
\epsfbox{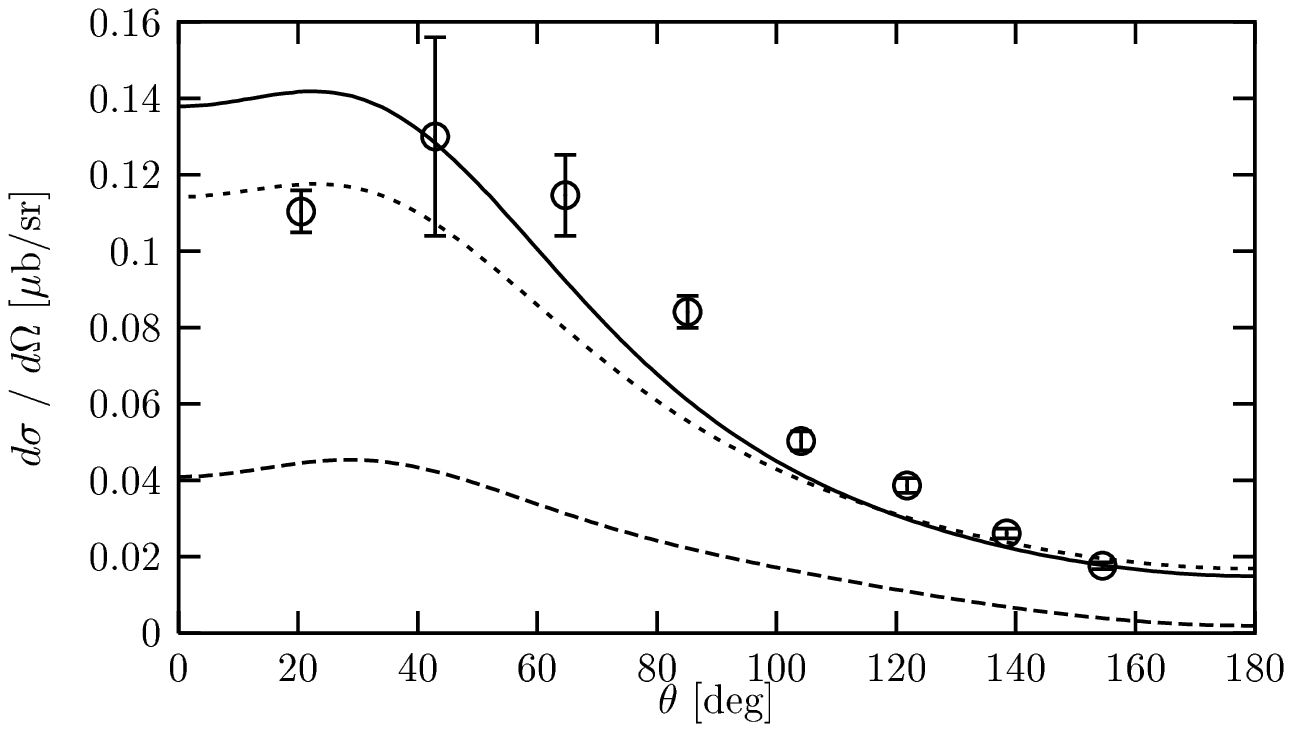}
\caption[]{The same as in Fig.~\protect\ref{fig75} at $E_p$= 200 MeV.
           $\theta$ is the c.m. $\gamma$-p scattering angle.
           Data are from~\protect\cite{Pickar}.}
\label{fig11}
\end{figure}

The photon angular distributions for pd capture are shown 
in Figs.~\ref{fig75}-\ref{fig11}.
The single nucleon current 
prediction is compared to the calculations including the Siegert
theorem, to the results adding the $\pi$- and $\rho$-like exchange currents
to the single nucleon current and to the data.
We see that the single nucleon current prediction underestimates 
the data. Siegert and explicit MEC's are close to the data 
and to each other, but still leaving a room for improvement. 
At the higher energies when $E_p$= 100, 150 and 200 MeV ($E_\gamma>$ 70 MeV) one
may note a slight superiority of the explicit MEC prediction over the Siegert 
approach.

\begin{figure}[hbt]
\epsfbox{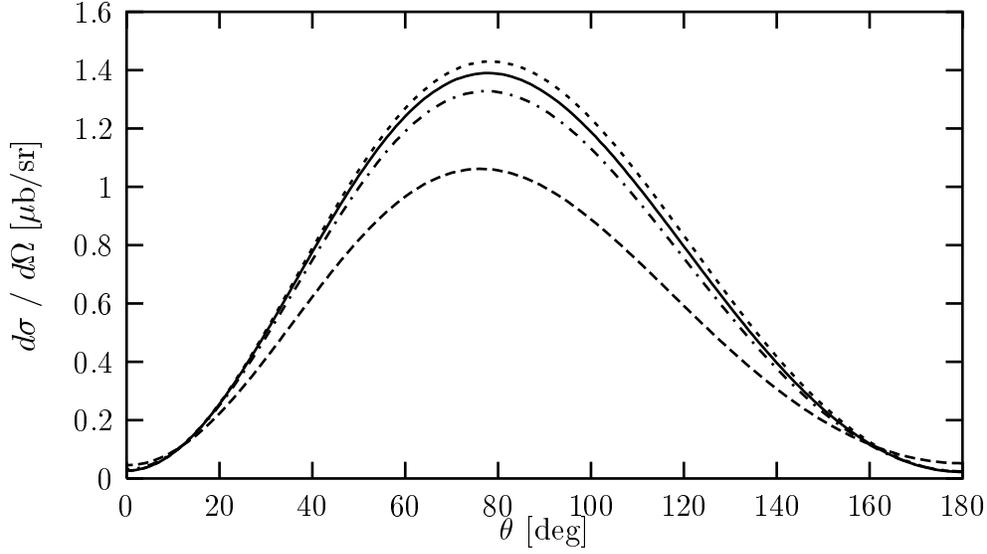}
\caption[]{The $^3$He angular distribution for pd capture at $E_d$= 17.5 MeV
           against the c.m. $^3$He-d scattering angle.
           The curves describe the single nucleon current (dashed),
           the single nucleon plus the $\pi$-MEC (dashed-dotted),
           the single nucleon plus the $\pi$- and $\rho$-MEC (solid)
           and Siegert (dotted) predictions.} 
\label{fig12}
\end{figure}

\begin{figure}[hbt]
\epsfbox{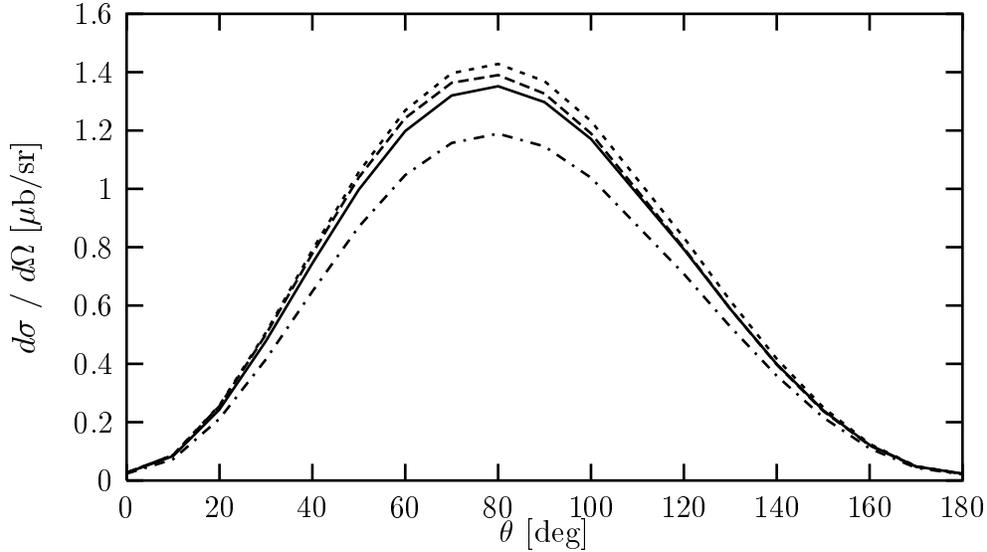}
\caption[]{The $^3$He angular distribution for pd capture at $E_d$= 17.5 MeV
           against the c.m. $^3$He-d scattering angle.
           The curves describe the Siegert without 3NF (dotted),
           the Siegert including 3NF (dashed-dotted),
           the single nucleon plus MEC without 3NF (dashed)
           and the single nucleon plus MEC including 3NF (solid) predictions.}
\label{fig13}
\end{figure}

\begin{figure}[hbt]
\epsfbox{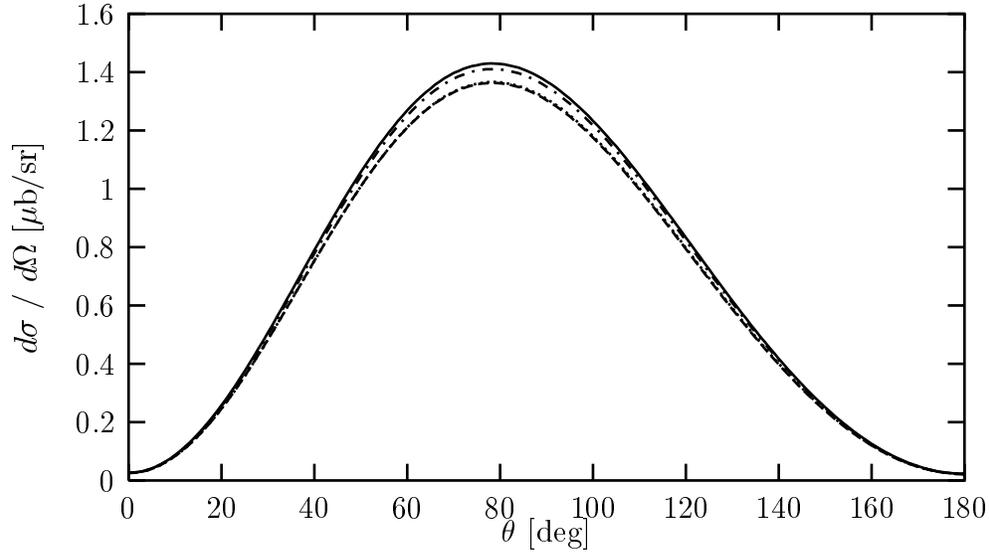}
\caption[]{The $^3$He angular distribution for pd capture at $E_d$= 17.5 MeV
           against the c.m. $^3$He-d scattering angle.
           The curves describe the Siegert predictions based on different
           NN forces:  Nijm~I (dotted), Nijm~II (dashed-dotted),
           CD Bonn (dashed) and AV18 (soild).}
\label{fig14}
\end{figure}

In Fig.~\ref{fig12} we show predictions for the $^3$He angular distribution
in the pd capture process for 
different choices of the electromagnetic current operator at a low energy. 
Apparently the Siegert choice is not fully exhausted by the electric multipole 
parts of the $\pi$- and $\rho$-like MEC's
but we see that the $\pi$-exchange MEC is the dominant one of the two. 
An interesting insight into 3NF effects is shown in Fig.~\ref{fig13}.
We use the Tuscon-Melbourne force~\cite{TM3NF},
which has been adjusted to the $^3$He binding energy~\cite{ref8}.
We see a relatively strong decrease of the peak height by including 
the 3NF in case of Siegert and a less pronounced shift in the same direction
in case of using MEC's. In \cite{Sandhas} it has been argued 
(see also \cite{Orlandini}) that this fact is related 
to the $^3$He binding energy. Apparently this can be more subtle
due to the interplay with the current operator used.
Finally we display the theoretical uncertainty arising from the fact that 
the various NN forces do not yield the same results (Fig.~\ref{fig14}).
This spread is relatively low ($\le$ 5 \% in the maximum), which 
is satisfying in case of Siegert. It will be very interesting to see in the future, 
whether different NN forces taken together with consistently applied MEC's will 
lead to a comparably small spread. Only if such a robustness can be
demonstrated one can have some confidence in this dynamical scenario.

\begin{figure}[hbt]
\epsfbox{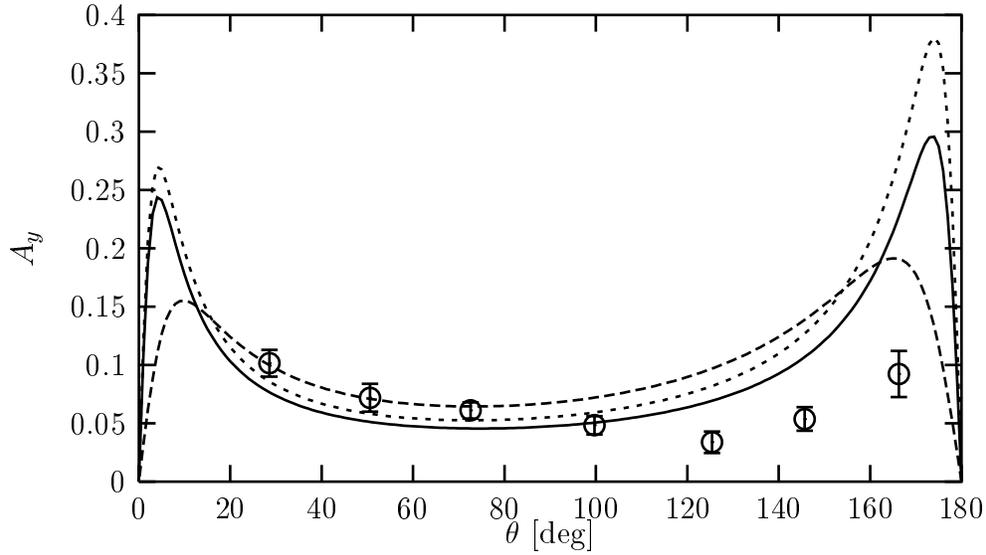}
\caption[]{The proton analyzing power $ A_y (p)$ at $E_p$= 5 MeV 
           against the c.m. $\gamma$-p scattering angle.
           Curves as in Fig.~\protect\ref{fig75}.
           Data are from~\protect\cite{Goeckner}.}
\label{fig15}
\end{figure}

\begin{figure}[hbt]
\epsfbox{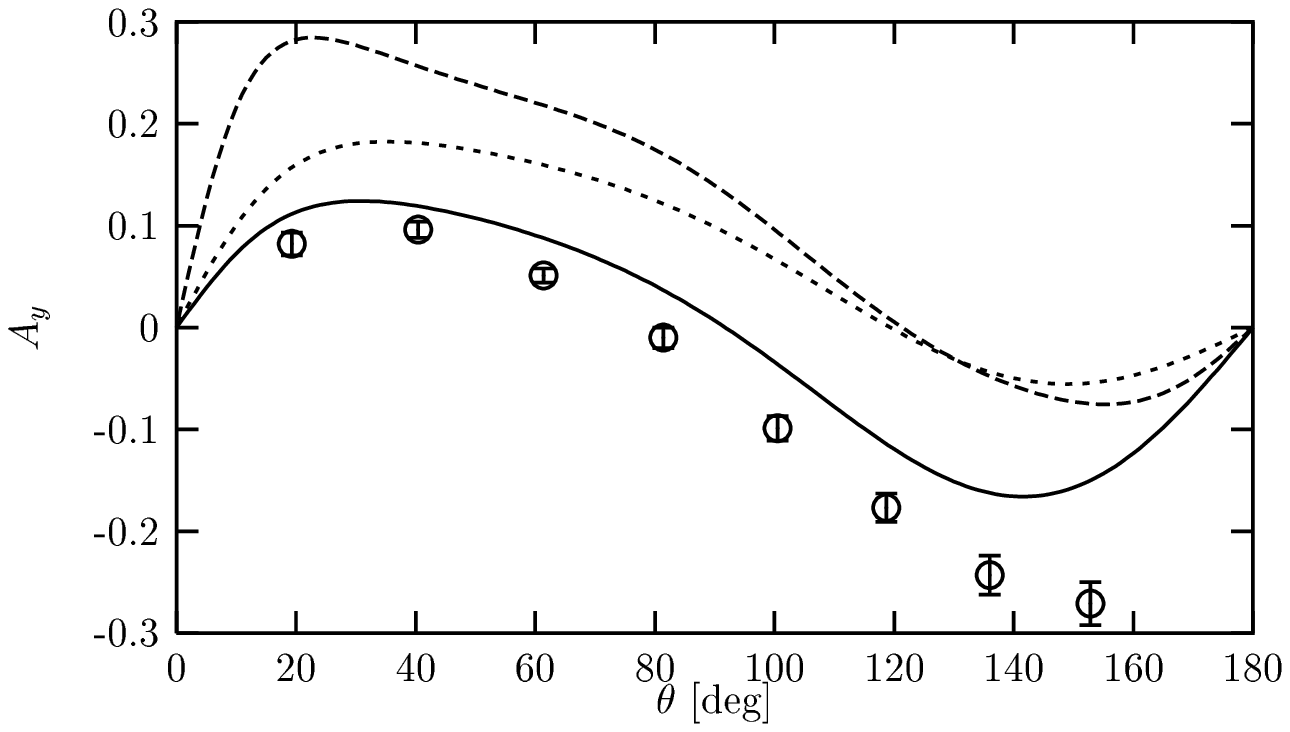}
\caption[]{The same as in Fig.~\protect\ref{fig15} at $E_p$= 100 MeV.
           Data are from~\protect\cite{Pickar}.}
\label{fig16}
\end{figure}

\begin{figure}[hbt]
\epsfbox{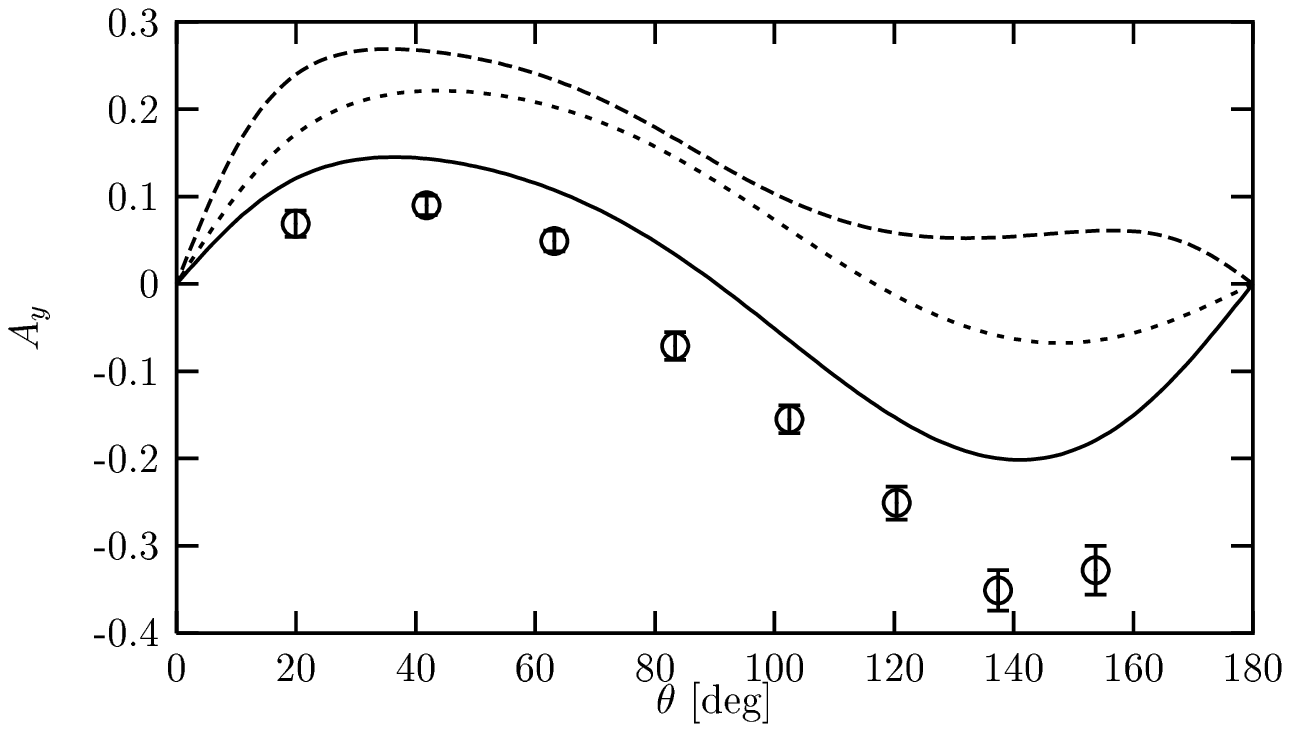}
\caption[]{The same as in Fig.~\protect\ref{fig15} at $E_p$= 150 MeV.
           Data are from~\protect\cite{Pickar}.}
\label{fig17}
\end{figure}

\begin{figure}[hbt]
\epsfbox{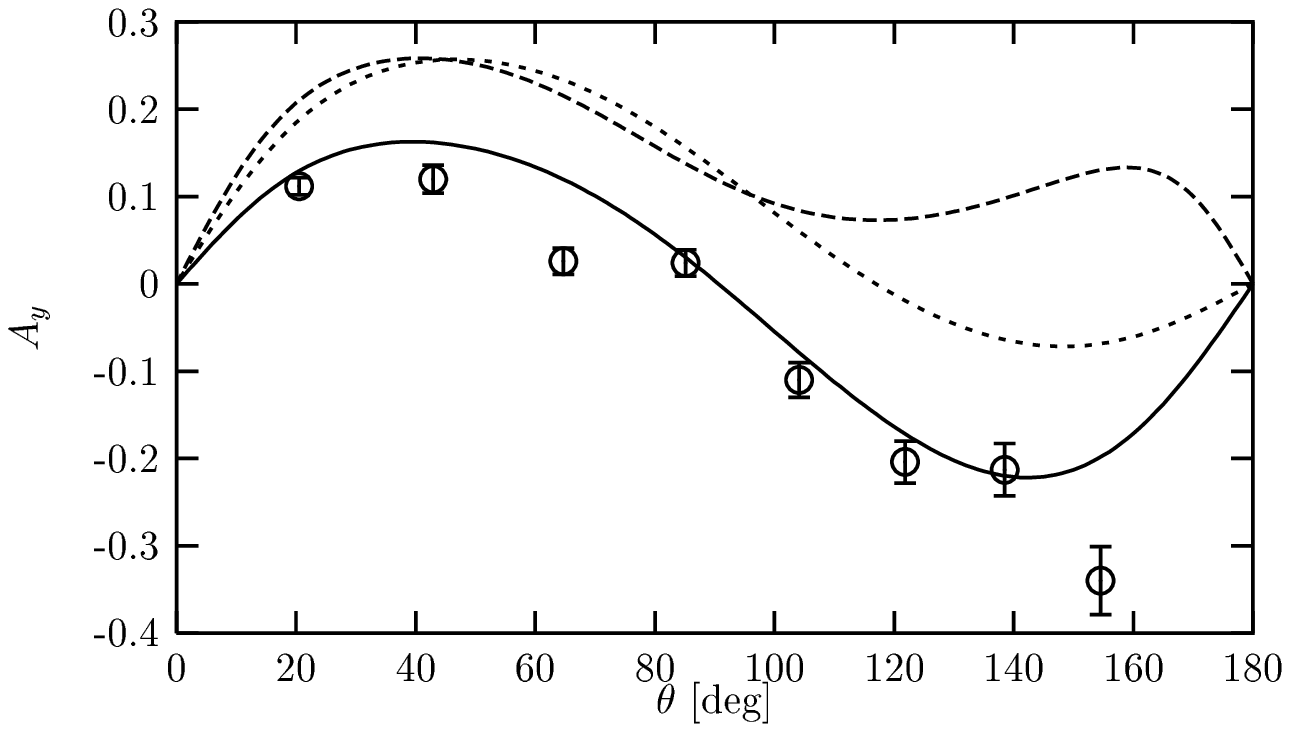}
\caption[]{The same as in Fig.~\protect\ref{fig15} at $E_p$= 200 MeV.
           Data are from~\protect\cite{Pickar}.}
\label{fig18}
\end{figure}

Let us now regard various spin observables. Nucleon analyzing powers $A_y (p)$ 
are shown in Figs.~\ref{fig15}--\ref{fig18}.
At the lowest energy of 5 MeV all three theoretical predictions are close 
together and near the data below 100 degrees, but at extreme 
forward and backward angles the two calculations including two-body currents 
show strong enhancements, which are not seen in the data at the backward 
angles. Very precise data especially at those extreme angles would be 
of interest to put higher pressure on the theory.
At the three higher energies the explicit use of MEC's is clearly much
closer to the data than the Siegert approach and clearly the single nucleon 
current alone is not acceptable.

\begin{figure}[hbt]
\epsfbox{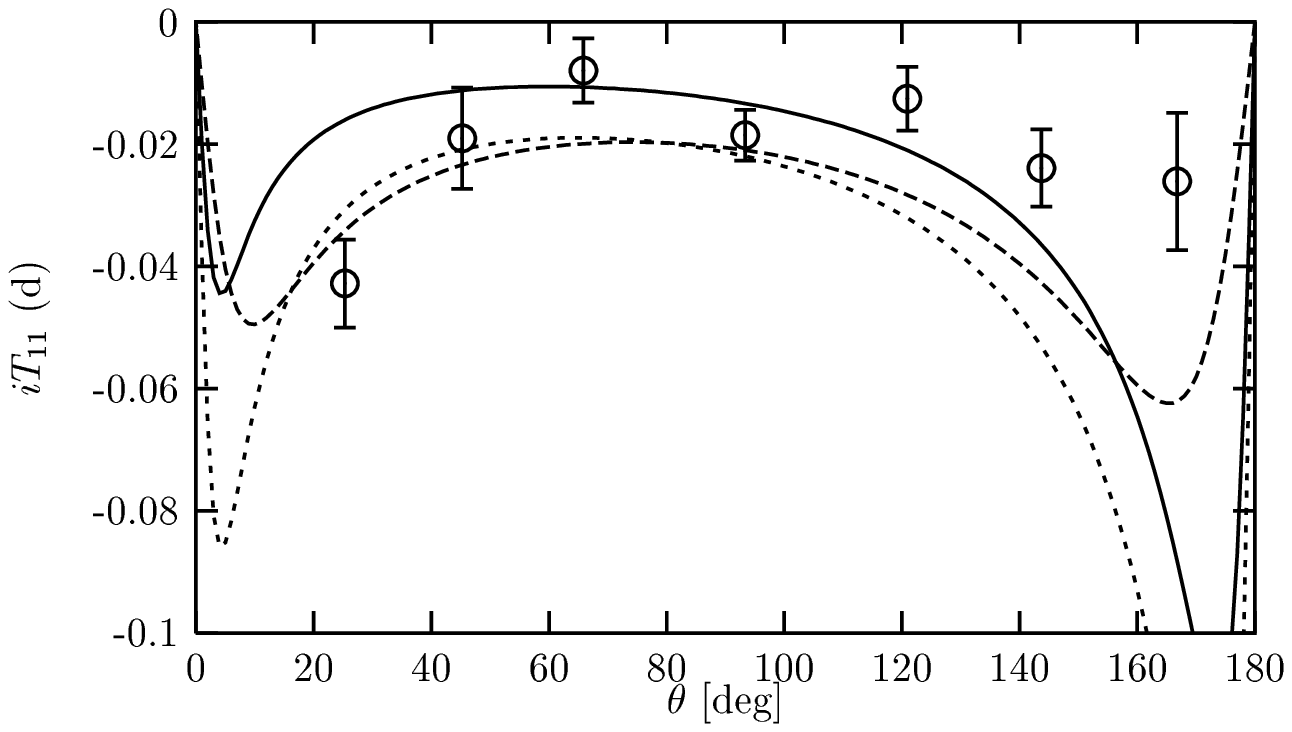}
\caption[]{The deuteron analyzing power $ i T_{11} $ at $E_d$= 10 MeV 
           against the c.m. $\gamma$-p scattering angle.
           Curves as in Fig.~\protect\ref{fig75}.
           Data are from~\protect\cite{Goeckner}.}
\label{fig19}
\end{figure}

\begin{figure}[hbt]
\epsfbox{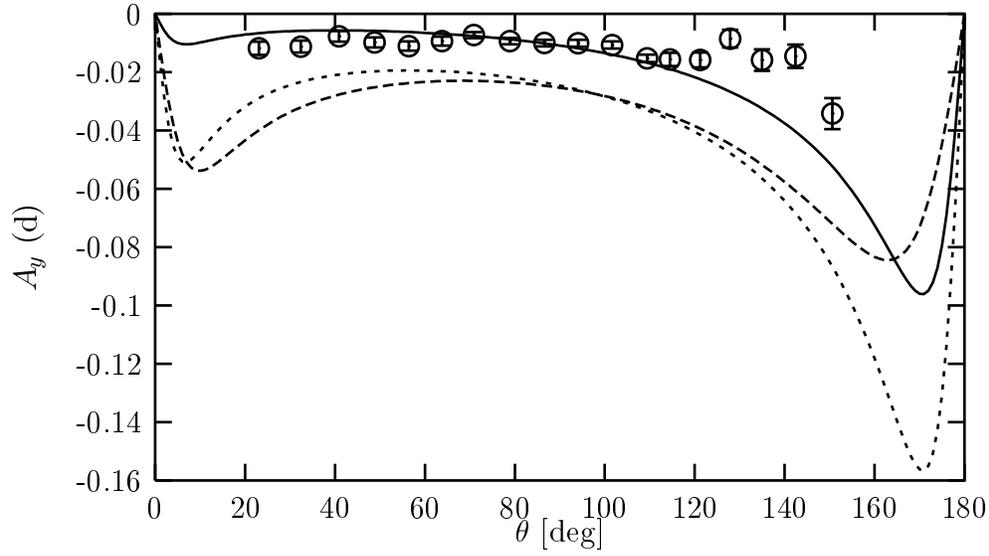}
\caption[]{The deuteron analyzing power $ A_{y} (d) $ at $E_d$= 17.5 MeV 
           against the c.m. $^3$He-d scattering angle.
           Curves as in Fig.~\protect\ref{fig75}.
           Data are from~\protect\cite{Sagara}.}
\label{fig20}
\end{figure}

\begin{figure}[hbt]
\epsfbox{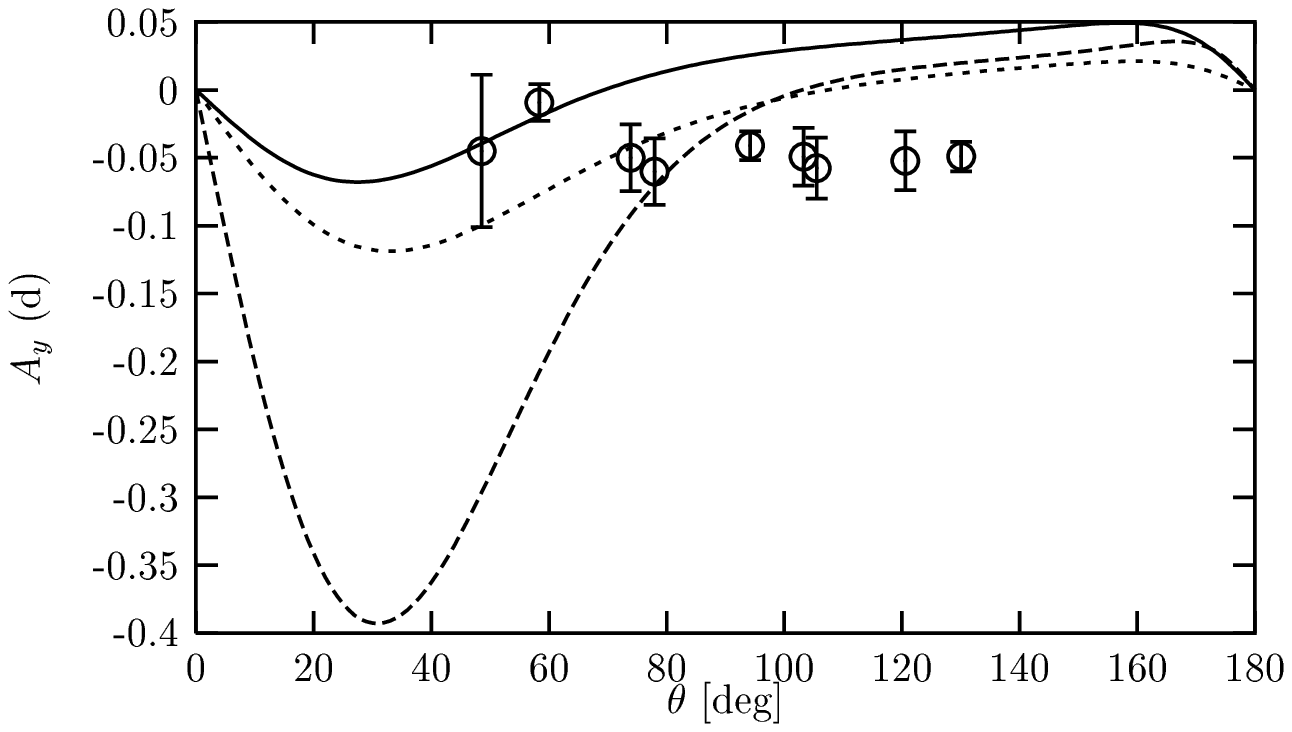}
\caption[]{The same as in Fig.~\protect\ref{fig20} at $E_d$= 95 MeV.
           $\theta$ is the c.m. $\gamma$-d scattering angle.
           Data are from~\protect\cite{Pitts}.}
\label{fig21}
\end{figure}

Our result for the deuteron vector analyzing power $iT_{11}$ or $A_y (d)$ are 
shown in Figs.~\ref{fig19}--\ref{fig21}. 
The agreement with the data is in general only fair.
At the lowest energy the extreme enhancement at large backward angles 
of the predictions including MEC or Siegert is clearly ruled out by the one
data point.
At $E_d$= 17.5 MeV the MEC prediction appears to be reasonable, if there 
would be not that enhancement at the large angles.
At $E_d$= 95 MeV Siegert and the explicit MEC
are comparable, but miss the data at backward angles. 

\begin{figure}[hbt]
\epsfbox{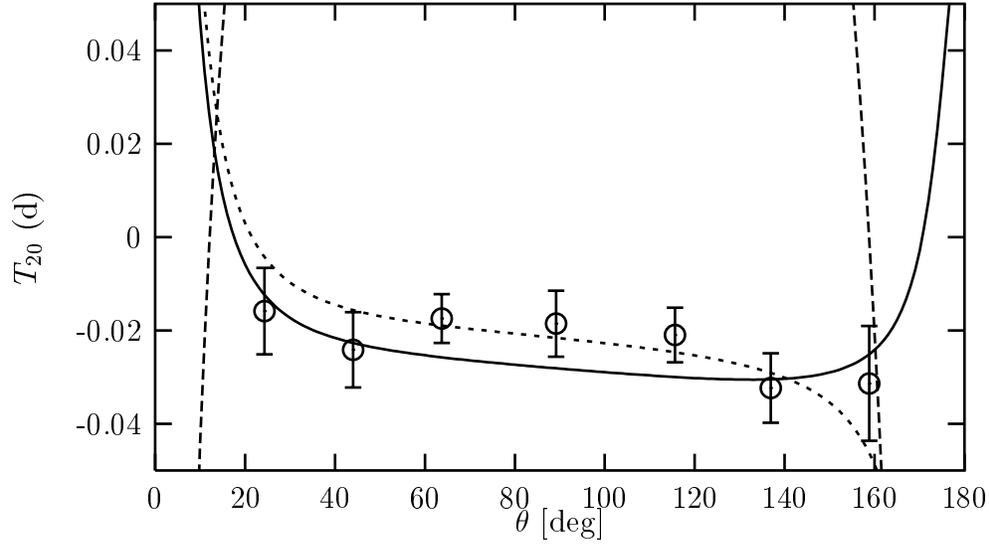}
\caption[]{The tensor analyzing power $ T_{20} $ at $E_d$= 10 MeV
           against the c.m. $\gamma$-p scattering angle.
           Curves as in Fig.~\protect\ref{fig75}.
           Data are from~\protect\cite{Goeckner}.}
\label{fig22}
\end{figure}

\begin{figure}[hbt]
\epsfbox{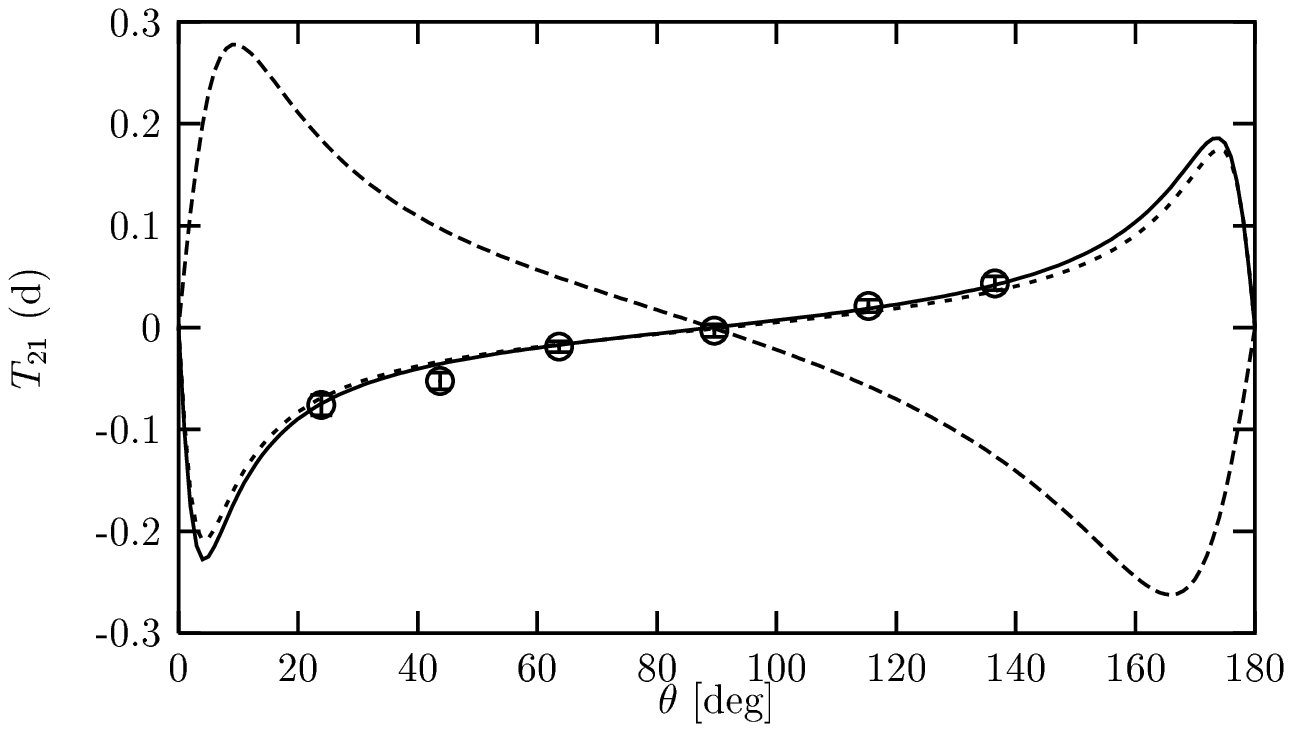}
\caption[]{The same as in Fig.~\protect\ref{fig22} for the tensor 
           analyzing power $ T_{21}$.
           Data are from~\protect\cite{Goeckner}.}
\label{fig23}
\end{figure}

\begin{figure}[hbt]
\epsfbox{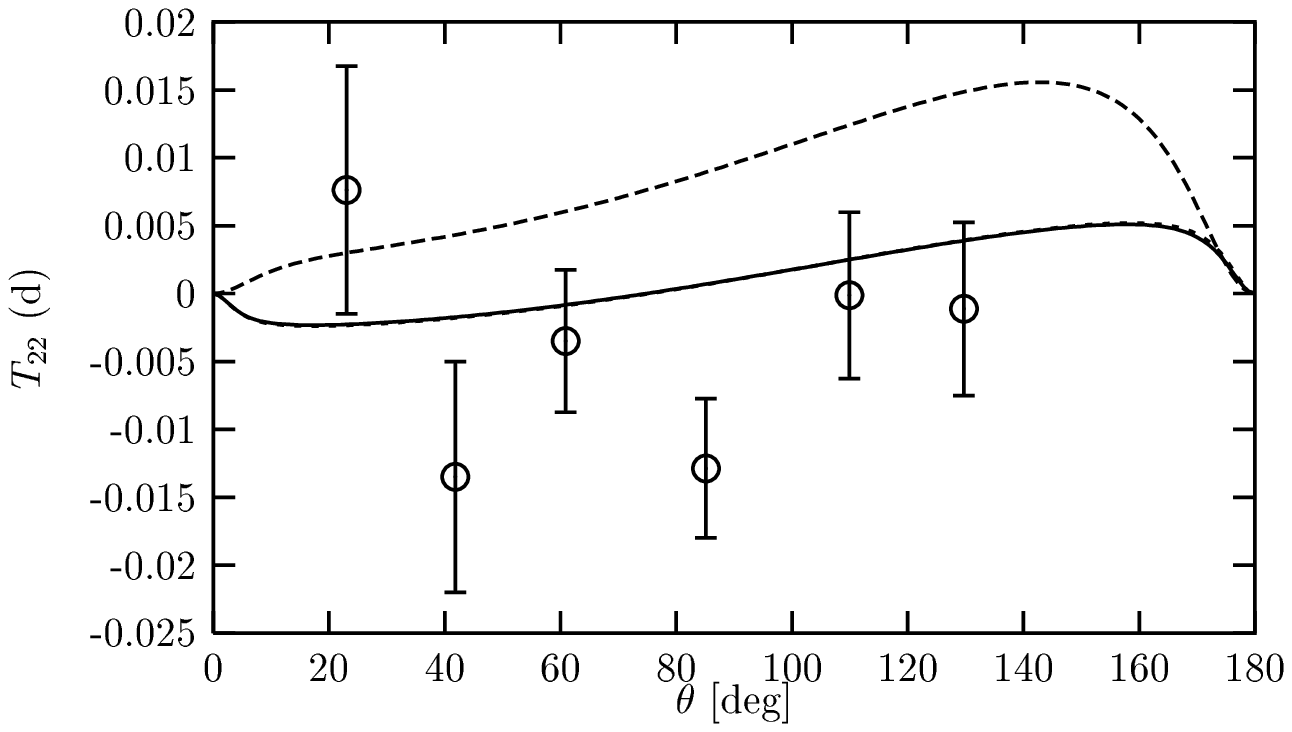}
\caption[]{The same as in Fig.~\protect\ref{fig22} for the tensor 
           analyzing power $ T_{22}$.
           Data are from~\protect\cite{Goeckner}.}
\label{fig24}
\end{figure}

\begin{figure}[hbt]
\epsfbox{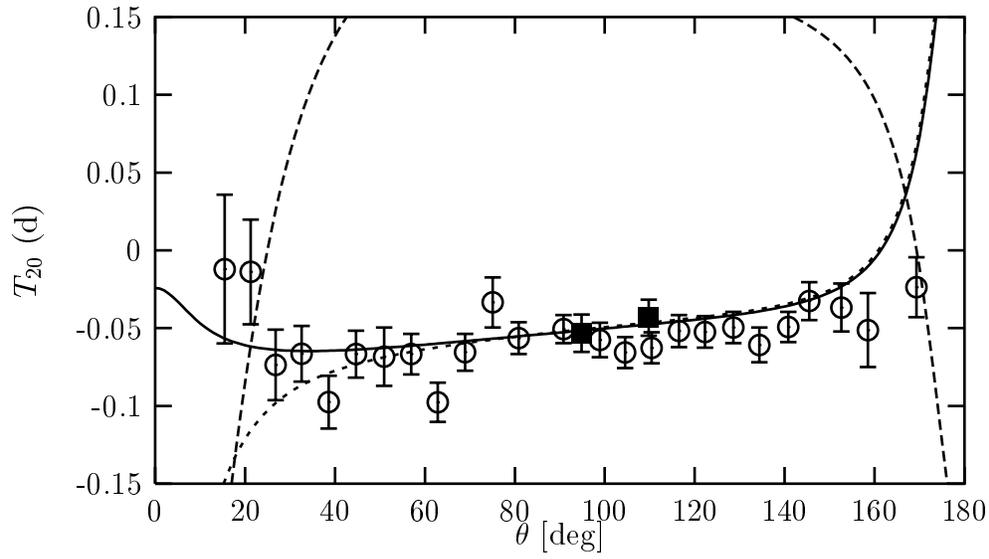}
\caption[]{The same as in Fig.~\protect\ref{fig22} at $E_d$= 19.8 MeV.
           Data are from~\protect\cite{Vetterli} (circles) and 
           from~\protect\cite{Schmid} (squares).}
\label{fig25}
\end{figure}

\begin{figure}[hbt]
\epsfbox{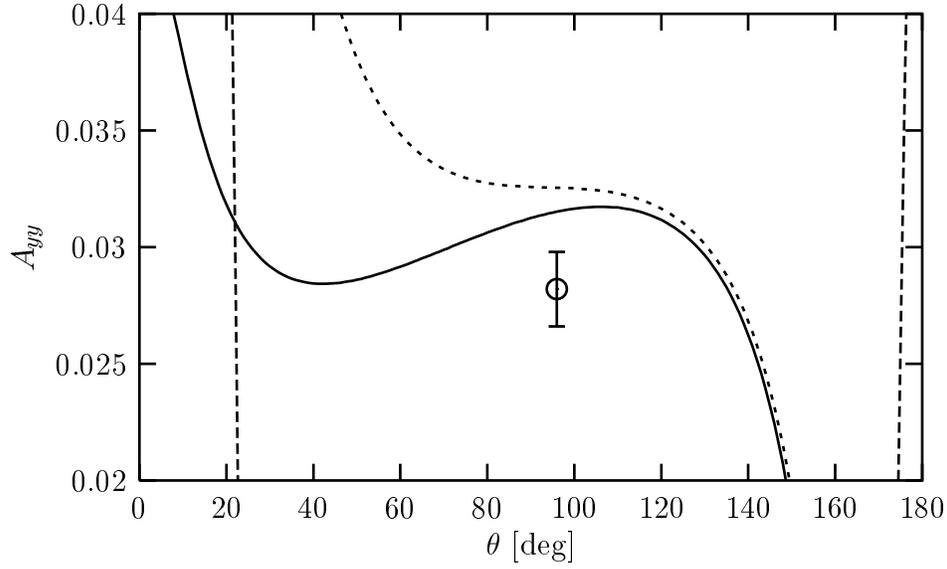}
\caption[]{The tensor analyzing power $ A_{yy} $ at $E_d$= 29.2 MeV
           against the c.m. $\gamma$-d scattering angle.
           Curves as in Fig.~\protect\ref{fig75}.
           The data point is from~\protect\cite{ref12}.}
\label{fig25.5}
\end{figure}

\begin{figure}[hbt]
\epsfbox{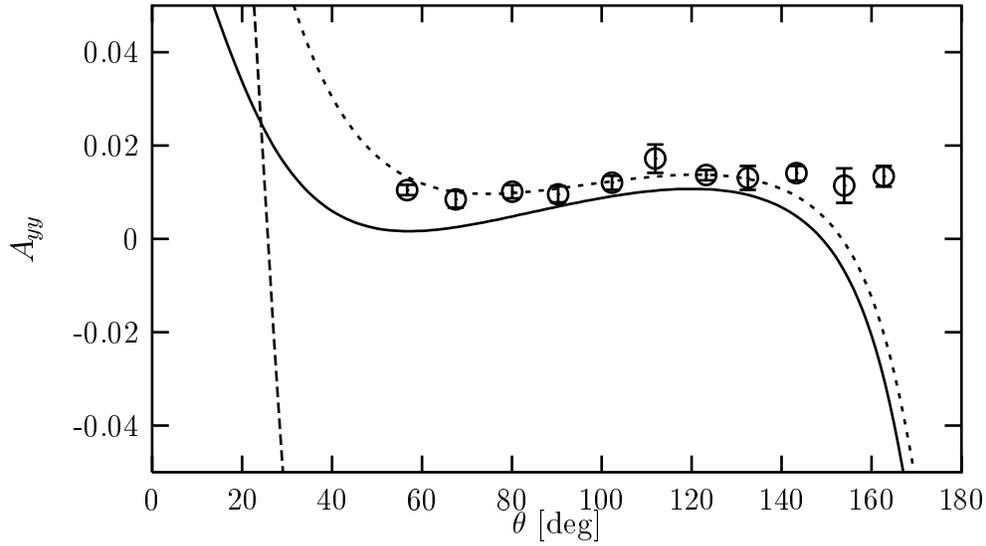}
\caption[]{The tensor analyzing power $ A_{yy} $ at $E_d$= 45 MeV
           against the c.m. $\gamma$-d scattering angle.
           Curves as in Fig.~\protect\ref{fig75}.
           Data are from~\protect\cite{ref20}.}
\label{fig26}
\end{figure}

\begin{figure}[hbt]
\epsfbox{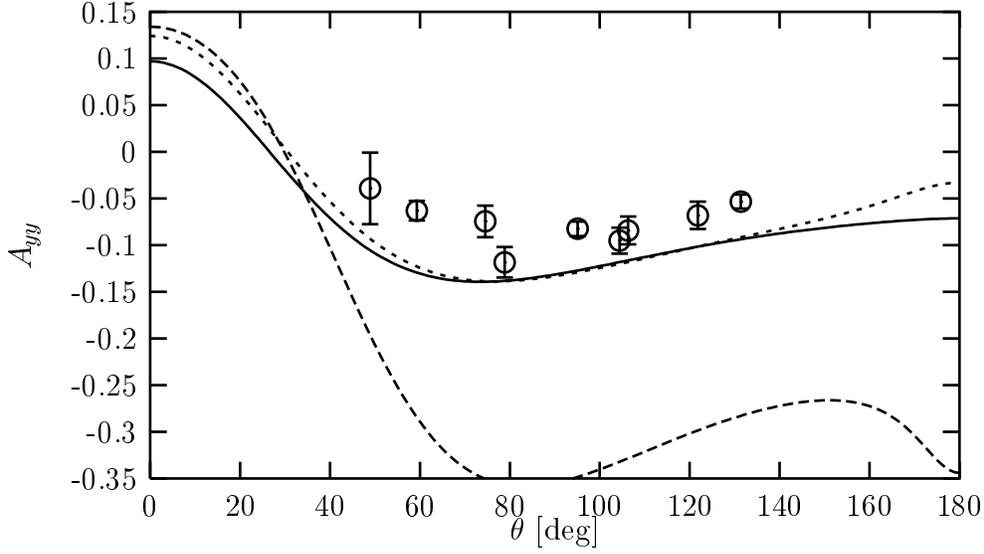}
\caption[]{The same as in Fig.~\protect\ref{fig26} at $E_d$= 95 MeV.
           Data are from~\protect\cite{Pitts}.}
\label{fig27}
\end{figure}

Last not least we regard a rich set of tensor analyzing powers 
in Figs.~\ref{fig22}--\ref{fig27}. At $E_d$= 10 MeV Siegert and MEC 
predictions agree with each other for $T_{20}$ and $T_{21}$ 
and the data. 
For the $T_{22}$ this is different and the data scatter a lot. We would
like to point to the very different behavior of the MEC and Siegert
predictions at large angles for $T_{20}$. The reason for that unacceptable
behavior is right now open. At $E_d$= 19.8 MeV this quite different behavior
for MEC and Siegert appears at extreme forward angles. Otherwise 
both predictions agree with each other and the data. 
At $E_d$= 29.2 MeV our results
(0.0326 for Siegert and 0.0315 for MEC)
for the single $A_{yy}$ data point ($\theta$= 96~$^\circ$) of~\cite{ref12}
agree with the calculations by Torre~\cite{ref12} (0.0339), although
in his case the Reid soft-core potential~\cite{Reid} was used.
$A_{yy}$ is fairly well described for $E_d$= 45 MeV
and $E_d$= 95 MeV. The data at $E_d$= 45 MeV have been 
analyzed by us before using the Bonn~B potential together with 
Siegert~\cite{ref20}.

\begin{figure}[hbt]
\epsfbox{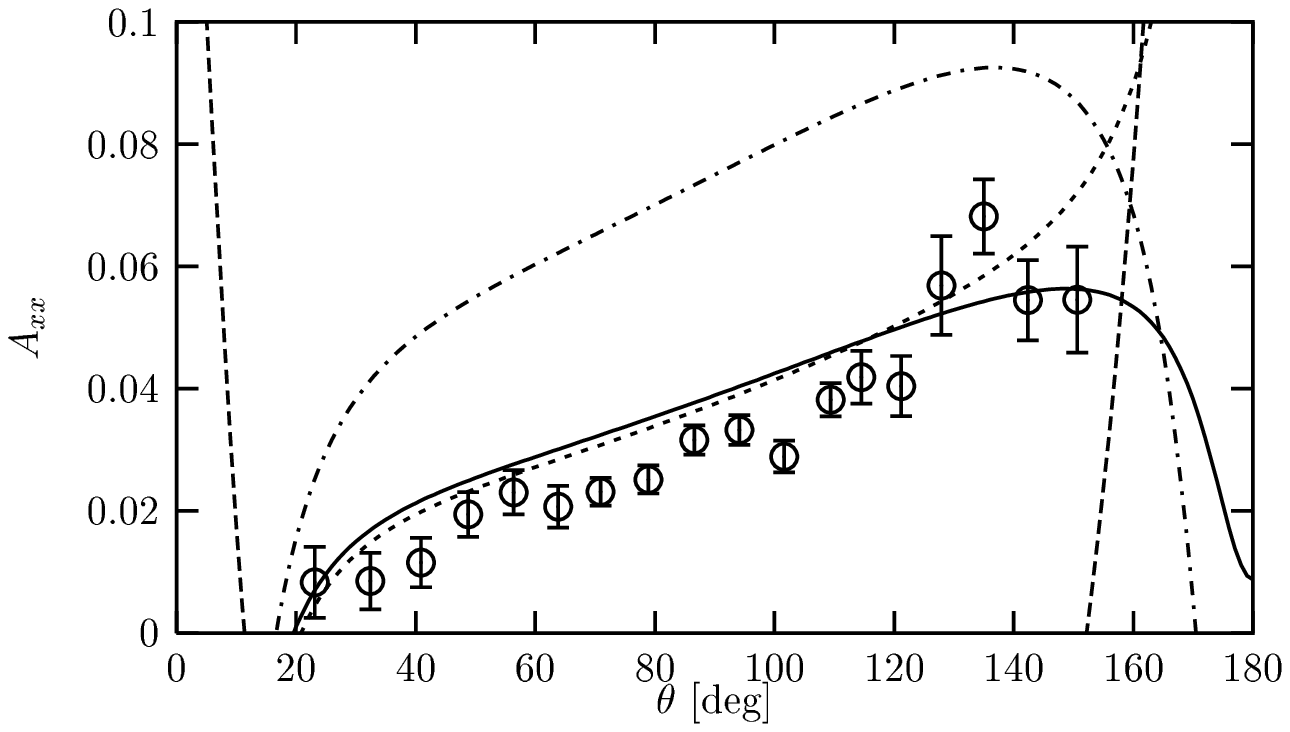}
\caption[]{The tensor analyzing power $ A_{xx} $ at $E_d$= 17.5 MeV
           against the c.m. $^3$He-d scattering angle.
           Curves as in Fig.~\protect\ref{fig12}.
           Data are from~\protect\cite{Sagara}.}
\label{fig28}
\end{figure}

\begin{figure}[hbt]
\epsfbox{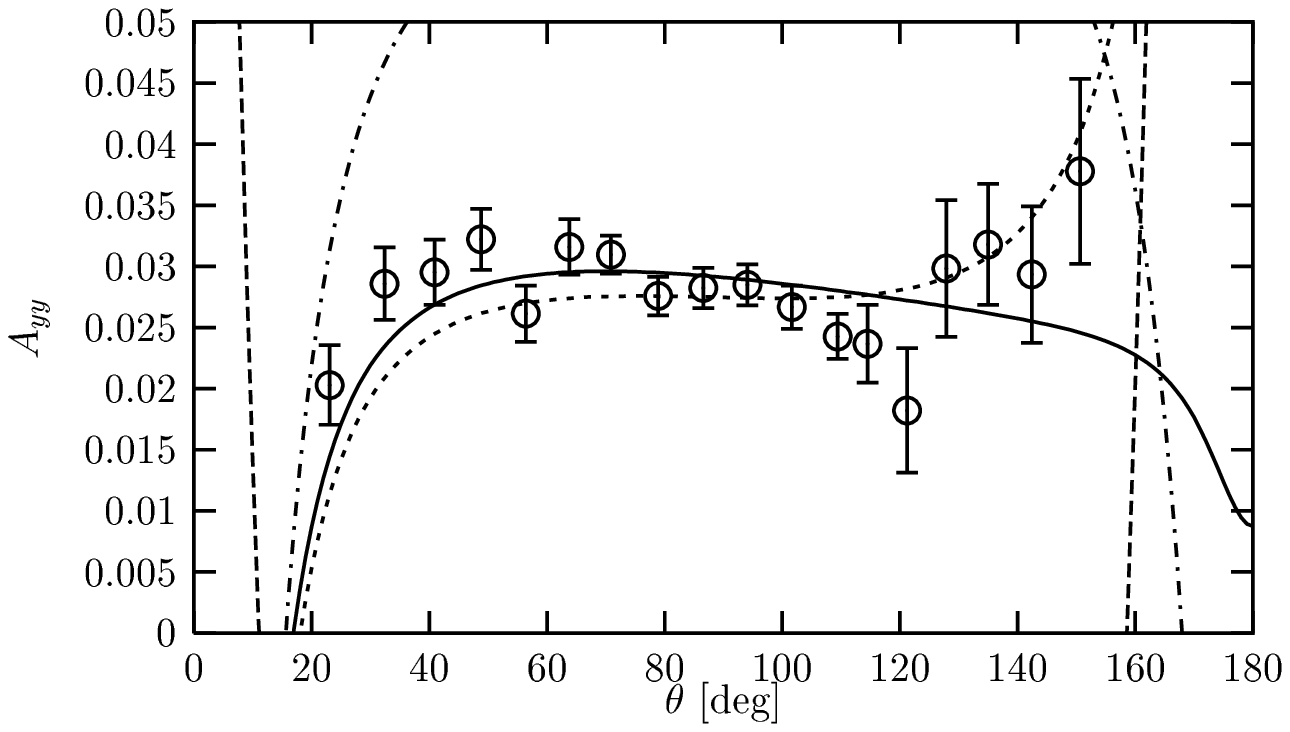}
\caption[]{The same as in Fig.~\protect\ref{fig28} for the 
           tensor analyzing power $ A_{yy}$.
           Data are from~\protect\cite{Sagara}.}
\label{fig29}
\end{figure}

\begin{figure}[hbt]
\epsfbox{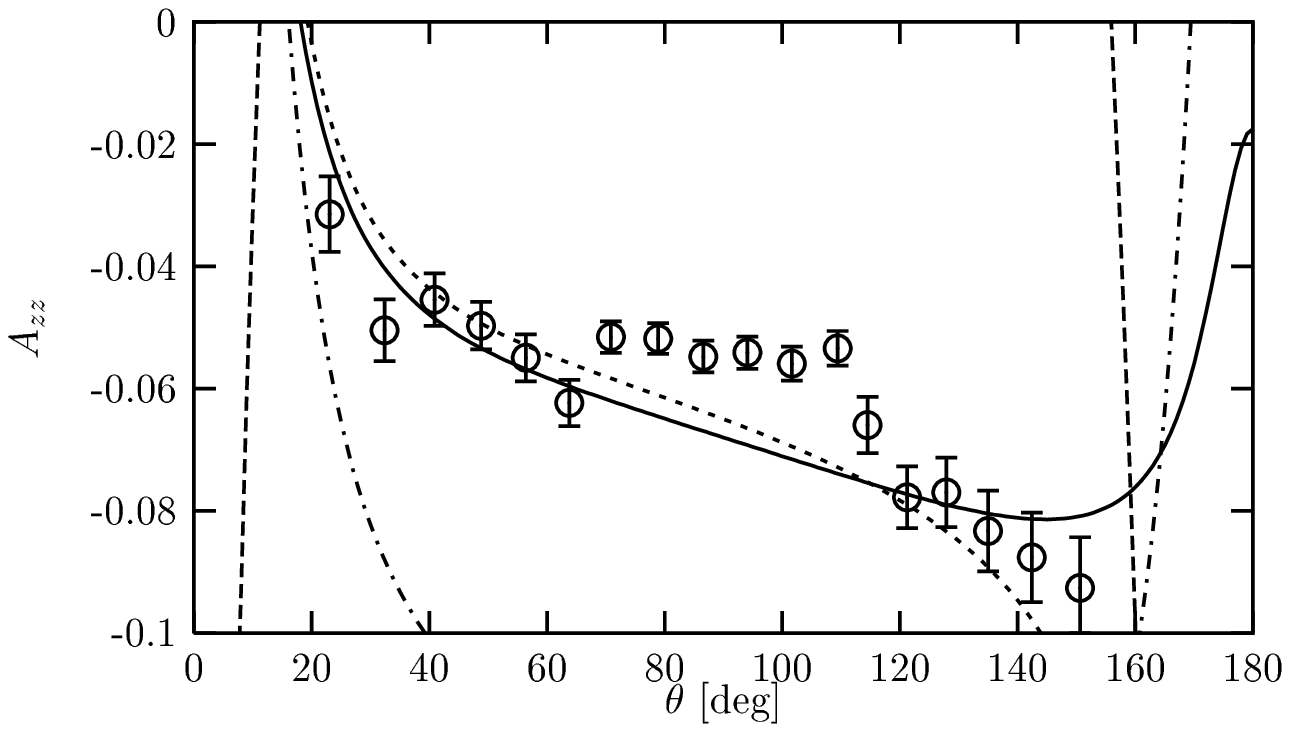}
\caption[]{The same as in Fig.~\protect\ref{fig28} for the 
           tensor analyzing power $ A_{zz}$.
           Data are from~\protect\cite{Sagara}.}
\label{fig30}
\end{figure}

At $E_d$= 17.5 MeV we show again in detail the various 
predictions. From Figs.~\ref{fig28}--\ref{fig30} one can infer that the 
$\pi$-exchange provides the strongest shift in relation to the single
nucleon current prediction, but the additional 
$\rho$-exchange is needed to bring the theory into the data. Also Siegert gives 
essentially the same quality of agreement. Remarkable are again the strong 
enhancements at extreme forward and backward
angles which unfortunately cannot be checked by available data. 

\begin{figure}[hbt]
\epsfbox{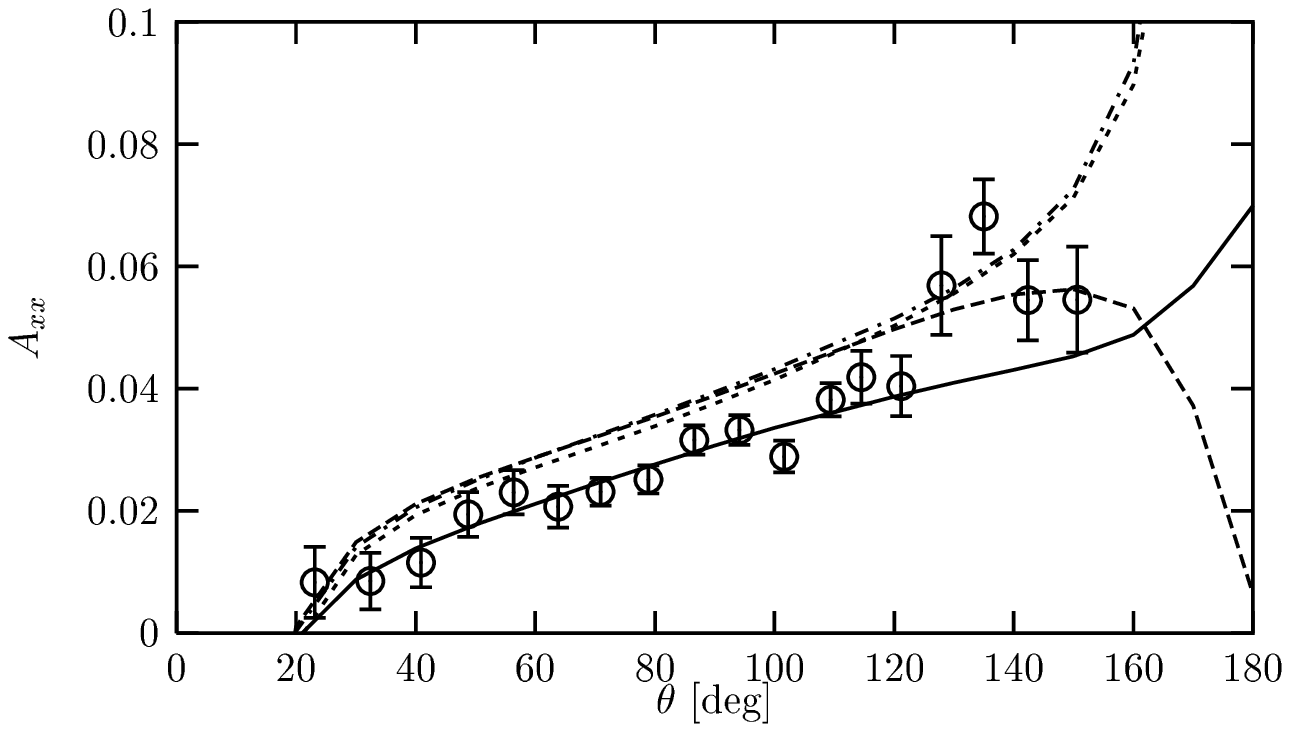}
\caption[]{The tensor analyzing power $ A_{xx} $ at $E_d$= 17.5 MeV
           against the c.m. $^3$He-d scattering angle.
           Curves as in Fig.~\protect\ref{fig13}.
           Data are from~\protect\cite{Sagara}.}
\label{fig31}
\end{figure}

\begin{figure}[hbt]
\epsfbox{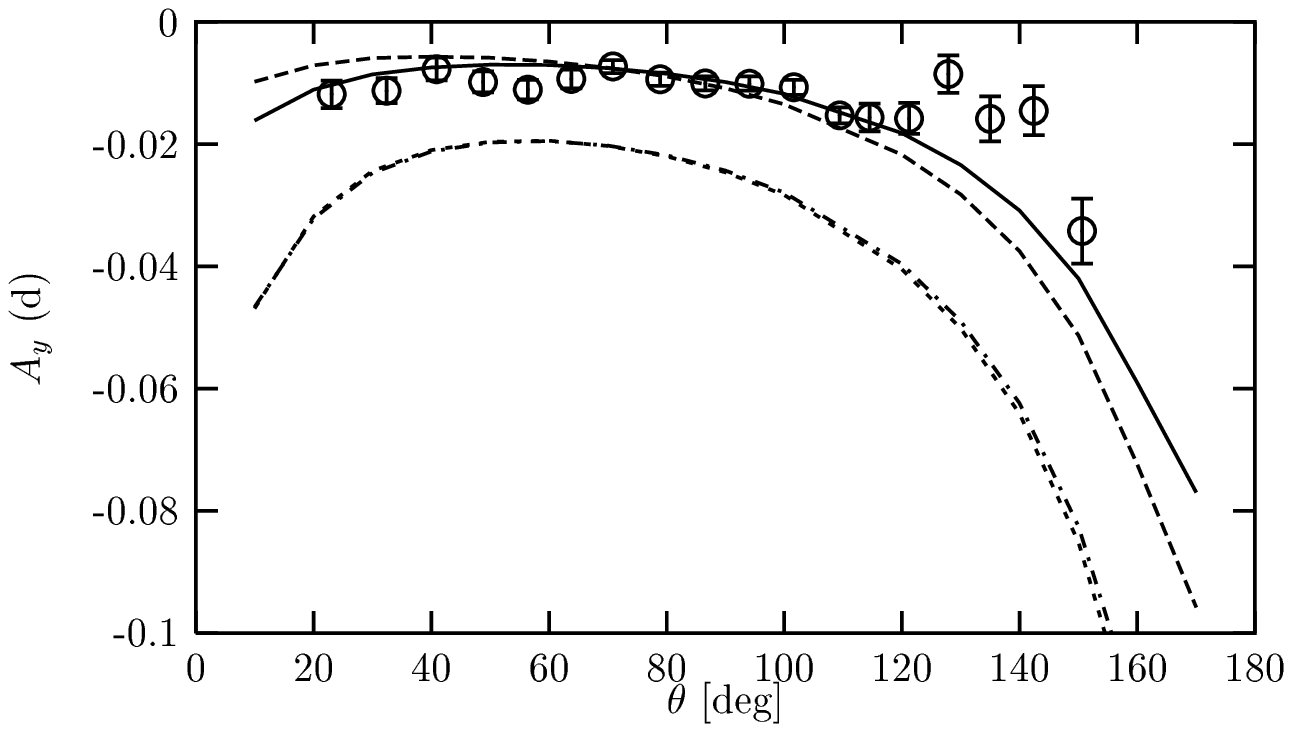}
\caption[]{The same as in Fig.~\protect\ref{fig31} for the deuteron
           vector analyzing power $ A_{y} (d)$.
           Data are from~\protect\cite{Sagara}.}
\label{fig32}
\end{figure}

\begin{figure}[hbt]
\epsfbox{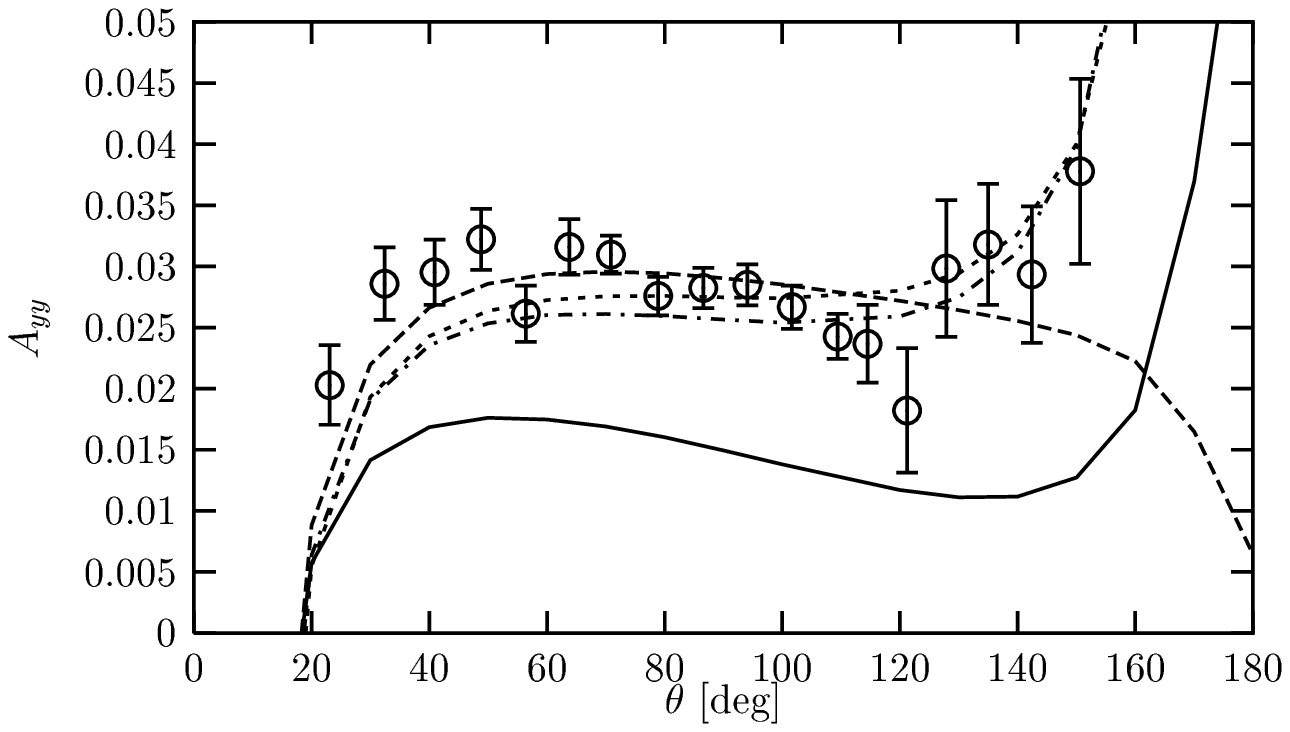}
\caption[]{The same as in Fig.~\protect\ref{fig31} for the 
           tensor analyzing power $ A_{yy}$.
           Data are from~\protect\cite{Sagara}.}
\label{fig33}
\end{figure}

\begin{figure}[hbt]
\epsfbox{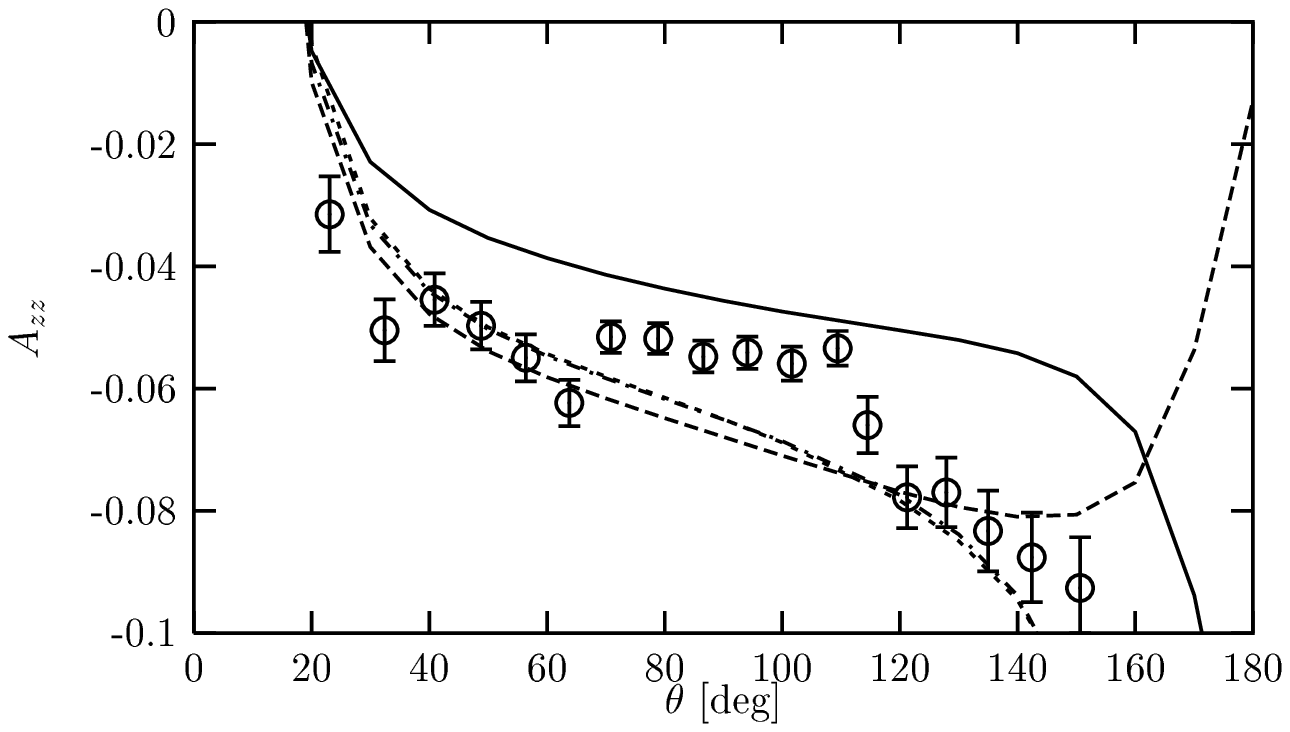}
\caption[]{The same as in Fig.~\protect\ref{fig31} for the 
           tensor analyzing power $ A_{zz}$.
           Data are from~\protect\cite{Sagara}.}
\label{fig34}
\end{figure}

We would like to add the remark that even at $E_d$= 17.5 MeV all states 
for two-nucleon total angular momenta up to at least j=2 have to be included.
All our calculations are based on up to j=3 contributions, which 
at the higher energies might be not fully converged.
In Figs.~\ref{fig31}--\ref{fig34} we display the effect of adding 
the Tuscon-Melbourne 3NF (adjusted to the $^3$He binding energy). 
For all observables the effect is negligible if taken together with Siegert.
In case of MEC's, however, the shifts are quite noticeable and move the 
theory somewhat away from the data in case of $A_{yy}$ and $A_{zz}$.

\begin{figure}[hbt]
\epsfbox{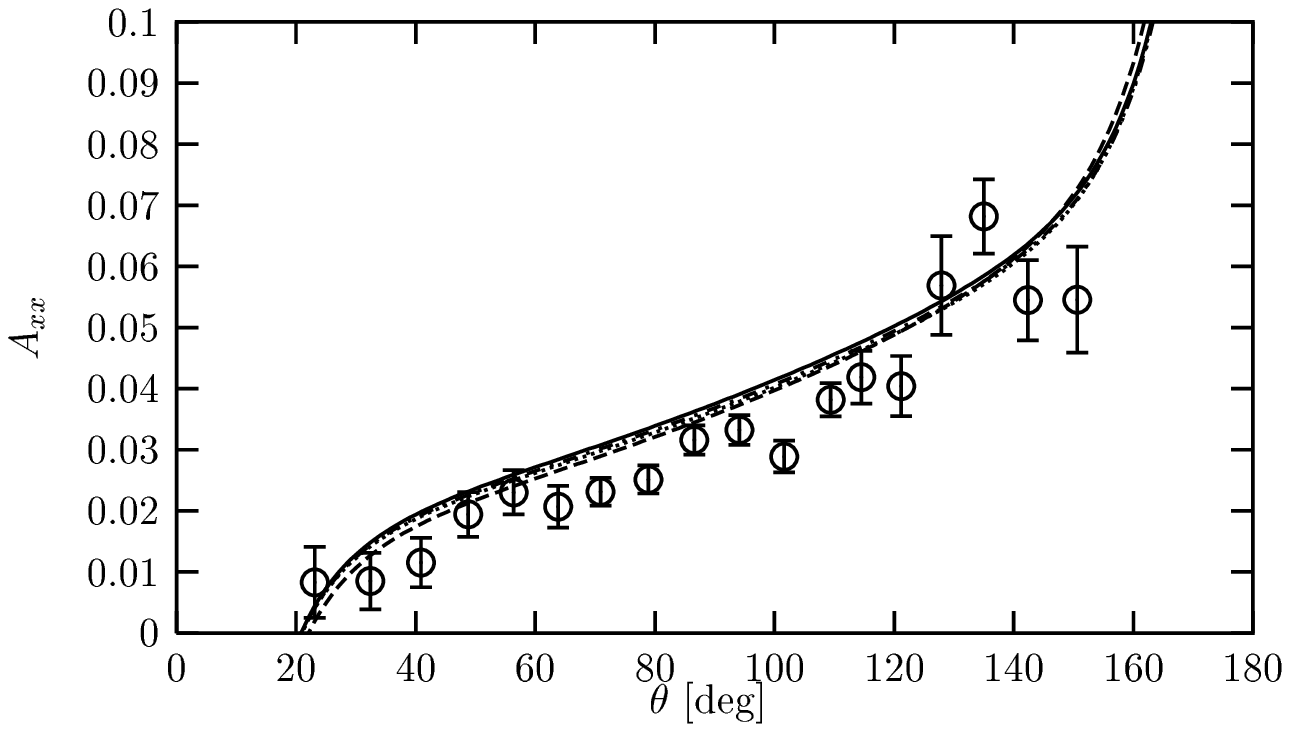}
\caption[]{The tensor analyzing power $ A_{xx} $ at $E_d$= 17.5 MeV
           against the c.m. $^3$He-d scattering angle.
           Curves as in Fig.~\protect\ref{fig14}.
           Data are from~\protect\cite{Sagara}.}
\label{fig35}
\end{figure}

\begin{figure}[hbt]
\epsfbox{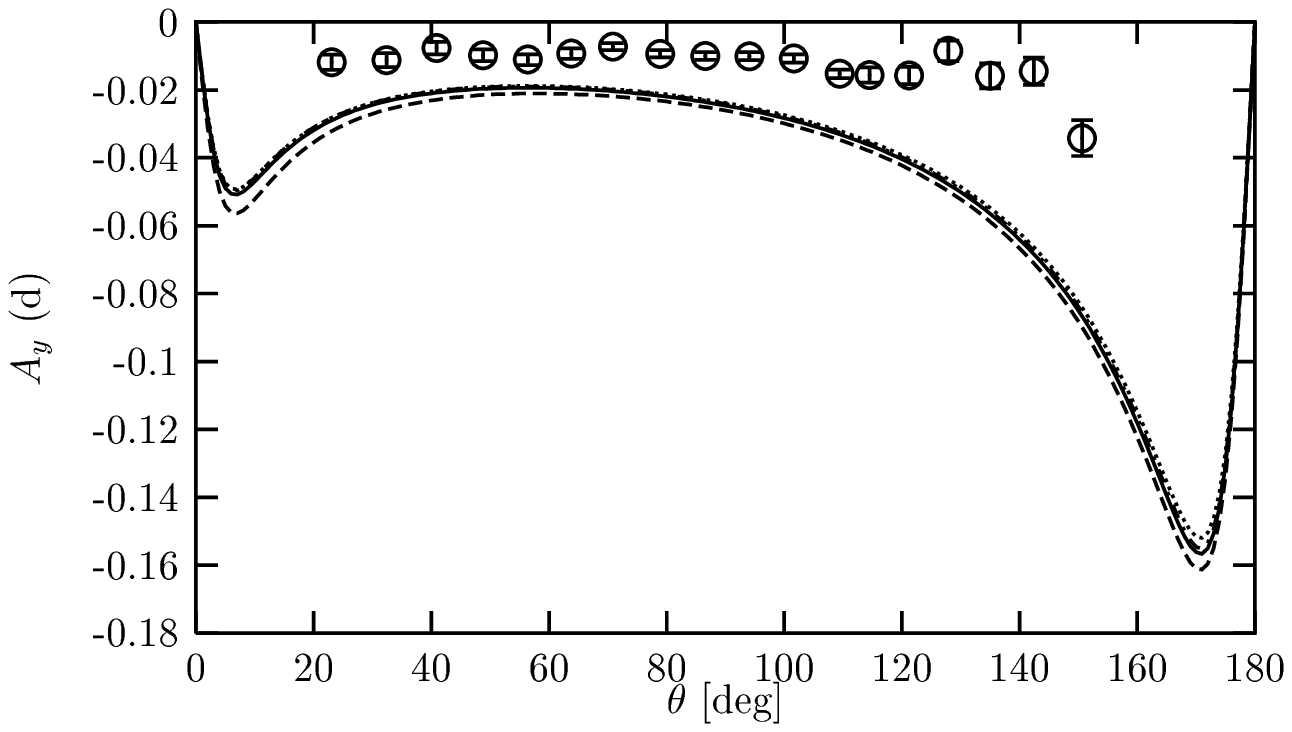}
\caption[]{The same as in Fig.~\protect\ref{fig35} for the deuteron
           vector analyzing power $ A_{y} (d)$.
           Data are from~\protect\cite{Sagara}.}
\label{fig36}
\end{figure}

\begin{figure}[hbt]
\epsfbox{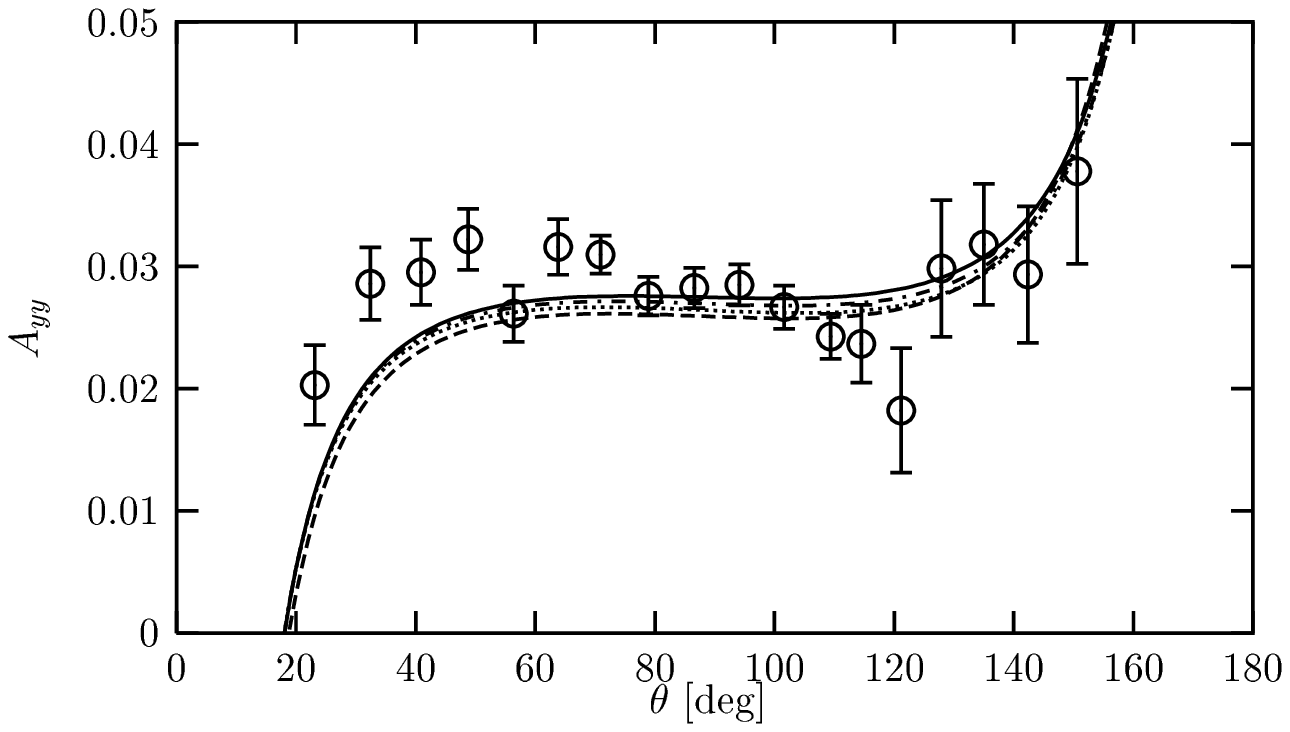}
\caption[]{The same as in Fig.~\protect\ref{fig35} for the 
           tensor analyzing power $ A_{yy}$.
           Data are from~\protect\cite{Sagara}.}
\label{fig37}
\end{figure}

\begin{figure}[hbt]
\epsfbox{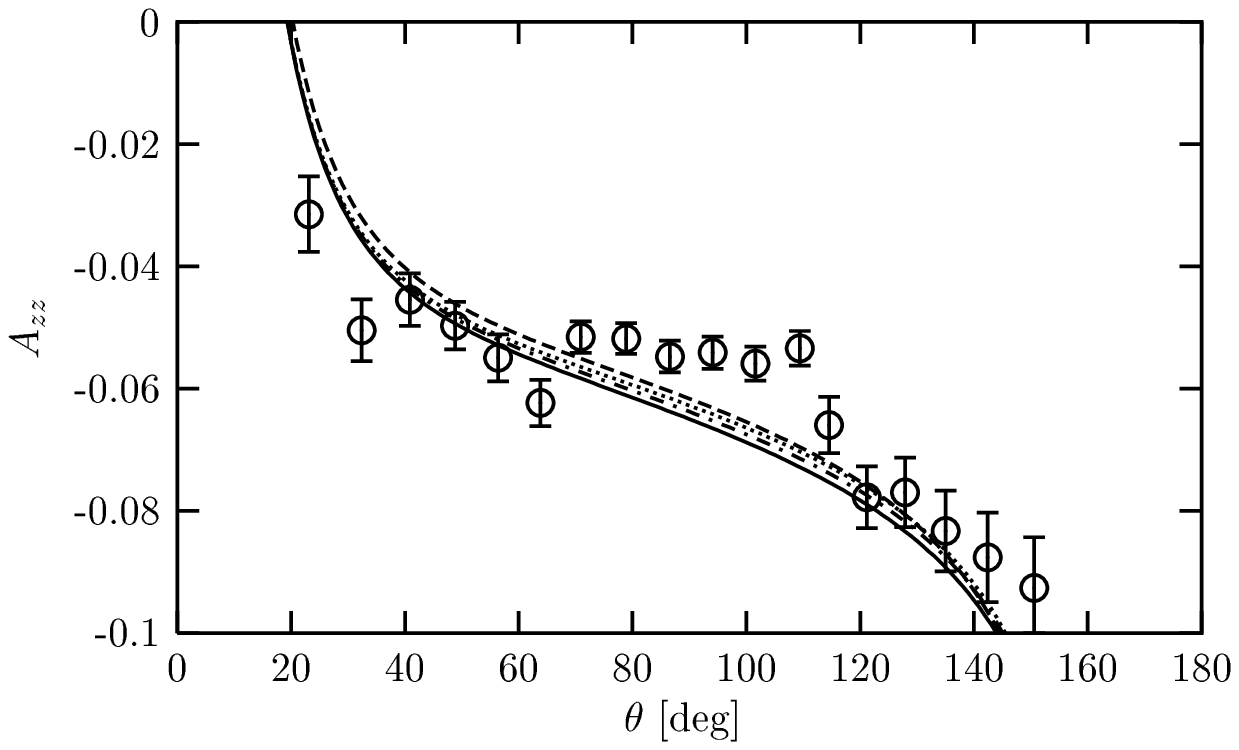}
\caption[]{The same as in Fig.~\protect\ref{fig35} for the 
           tensor analyzing power $ A_{zz}$.
           Data are from~\protect\cite{Sagara}.}
\label{fig38}
\end{figure}

Finally we demonstrate in Figs.~\ref{fig35}--\ref{fig38}
that the different NN forces taken together with Siegert give
essentially the same predictions, which is a very nice feature of 
robustness of that dynamical picture.

\section{Summary}
\label{secV}

We have analyzed the angular distributions and some polarization observables
in the proton--deuteron radiative capture at low and intermediate energies. 
The corresponding photon lab energy for the inverse photodisintegration 
process would range from 10 to 140 MeV.
We compared the explicit use of MEC's to the use of Siegert's theorem.
The MEC's were restricted 
to $\pi$- and $\rho$-like exchanges as derived from AV18 according to Riska's 
prescription. The 3N bound and scattering states are rigorous solutions
of the adequate Faddeev equations. The calculations are practically
converged with respect to angular momentum states except possibly at the higher
energies. At the lower energies the Siegert
and MEC predictions are rather close together. The $\pi$-like
MEC provides the strongest shift in relation to the single nucleon current
towards the Siegert's result, but the $\rho$-like piece is important
as another shift. At the higher energies Siegert and MEC's differ
in general, which is to be expected since the Siegert approach 
is mainly active for the lower multipoles. 

The agreement with the data 
is mostly good but there is some room for improvement. 
Definitely new measurements are needed to improve on certain data sets 
and to put stronger constraints on the behavior of theoretical predictions
at extreme angles.
Needless to say that
the correct treatment of the initial state interaction is required
in all cases we studied.
PWIA results are very poor (we did not even display them).

The strong and sometimes opposite enhancements of the MEC and Siegert 
predictions at extreme angles are worthwhile to be mentioned.
Precise data would be very useful to rule that out or possibly
justify it. In any case if the MEC and the Siegert predictions 
are opposite, the reason for that should be clarified.

We consider it to be an important result that the different new generation
NN force 
predictions together with Siegert are very close together, which demonstrates
the stability of that dynamical picture. It will be important to investigate
in future studies, whether this is also true, when explicit MEC's are used
consistently to the NN forces. Here we restricted ourselves to just one case,
when working with MEC, the AV18 potential.

Finally we mention that the inclusion of 3NF's in the form of 
the Tuscon-Melbourne $2 \pi$-exchange (adjusted to the $^3$He binding energy)
has only a minor effect in conjunction with Siegert, but a noticeable one
(in some cases) if used together with MEC's. Clearly consistency 
requirements will probably play an essential role and are not the subject
of this article.

\acknowledgements

This work was supported by
the Deutsche Forschungsgemeinschaft (H.K. and J.G.),
the Polish Committee for Scientific Research under Grant No. 2P03B03914
and the Science and Technology Cooperation Germany-Poland. 
One of us (W.G.) would like to thank the Foundation for Polish Science
for the financial support during his stay in Cracow.
The numerical calculations have been performed on the Cray T90 of the
NIC in J\"ulich, Germany.

\appendix
\section{An alternative approach to the multipole decomposition}

Here we would like to present an independent derivation of Eq.~(\ref{eq:eq6}),
which differs substantially in technical tools from the one given 
in section~\ref{secII}.

The nuclear matrix elements ${\vec I \left(\vec Q\right)}$ can be expanded 
in a series of vector
spherical harmonics
\begin{equation}\label{IIY}
{\vec I \left(\vec Q\right)} =\sum_{lJ\xi}I^{\xi}_{Jl}(Q)\;{\vec Y^{\:*{\xi}}_{Jl1} {\left(\hat Q\right)}},
\end{equation}
where
\begin{equation}\label{IMJLQ}
I^{\xi}_{Jl}(Q)=\int d{\hat Q\,'}{\vec I \left(\vec Q\,'\right)} 
 {\vec Y^{\:{\xi}}_{Jl1} {\left(\hat Q\,'\right)} }.
\end{equation}
We denote $Q=\left|{\vec Q}\right|=\left| {\vec Q\,'} \right|.$
Since $l$ may take values $l=|J-1|,J,J+1$ in Eq.~(\ref{IIY}), we have
\begin{equation}
{\vec I \left(\vec Q\right)}  =\sum_{J{\xi}} \left(
I^{\xi}_{JJ-1}(Q)\;{\vec Y^{\:*{\xi}}_{JJ-11}{\left(\hat Q\right)}} \, + \,
I^{\xi}_{JJ}(Q)\;{\vec Y^{\:*{\xi}}_{JJ1} {\left(\hat Q\right)} } \, + \,
I^{\xi}_{JJ+1}(Q)\;{\vec Y^{\:*{\xi}}_{JJ+11} {\left(\hat Q\right)} } \right) .
\label{sum3}
\end{equation}
We define
\begin{equation}
{\vec I^{\,el}_{J{\xi}} ({\vec Q} )} = I^{\xi}_{JJ-1}(Q)\;{\vec Y^{\:*{\xi}}_{JJ-11} {\left(\hat Q\right)}} \, + \,
I^{\xi}_{JJ+1}(Q)\;{\vec Y^{\:*{\xi}}_{JJ+11} {\left(\hat Q\right)} }
\label{el}
\end{equation}
and
\begin{equation}
{\vec I^{\,mag}_{J{\xi}} ({\vec Q})} = I^{\xi}_{JJ}(Q)\;{\vec Y^{\:*{\xi}}_{JJ1} {\left(\hat Q\right)} } .
\label{mag}
\end{equation}
In Eqs.~(\ref{el}) and (\ref{mag}) we separate terms of electric
and magnetic type
containing the vector spherical harmonics with parity $(-1)^J$
and $(-1)^{(J+1)}$ respectively
\begin{equation}\label{IIelImag}
{\vec I \left(\vec Q\right)}  =\sum_{J{\xi}}\left({\vec I^{\,el}_{J{\xi}} ({\vec Q} )}
+{\vec I^{\,mag}_{J{\xi}} ({\vec Q})} \right).
\end{equation}
Then we use the following relations~\cite{VMK88}
\begin{equation}
\hat{Q} \times  {\vec Y^{\:{\xi}}_{JJ+11} {\left(\hat Q\right)} } \ = \ 
i\, \sqrt{ { J \over {2J+1}} } \, {\vec Y^{\:{\xi}}_{JJ1}
{\left(\hat Q\right)} }
\label{id1}
\end{equation}
\begin{equation}
\hat{Q} \times  {\vec Y^{\:{\xi}}_{JJ1} {\left(\hat Q\right)}  } \ = \ 
i\, \sqrt{ { {J+1} \over {2J+1}} } \, {\vec Y^{\:{\xi}}_{JJ-11}
{\left(\hat Q\right)}   }
\ + \
i\, \sqrt{ { {J} \over {2J+1}} } \, {\vec Y^{\:{\xi}}_{JJ+11}
{\left(\hat Q\right)}  }
\label{id2}
\end{equation}
\begin{equation}
\hat{Q} \times  {\vec Y^{\:{\xi}}_{JJ-11} {\left(\hat Q\right)} } \ = \ 
i\, \sqrt{ { {J+1} \over {2J+1}} } \, {\vec Y^{\:{\xi}}_{JJ1}
{\left(\hat Q\right)}  }
\label{id3}
\end{equation}
\begin{equation}
\hat{Q}\, {Y_{J{\xi}}{\left(\hat Q\right)} } \ = \ 
\sqrt{ { {J} \over {2J+1}} } \, {\vec Y^{\:{\xi}}_{JJ-11}
{\left(\hat Q\right)} }
\ - \ \sqrt{ { {J+1} \over {2J+1}} } \, {\vec Y^{\:{\xi}}_{JJ+11}
{\left(\hat Q\right)}  }
\label{id35}
\end{equation}
to derive the following identities
\begin{equation}
{\vec Y^{\:{\xi}}_{JJ+11} {\left(\hat Q\right)}  } \ = \ 
- \sqrt{ { {J+1} \over {2J+1}} } \, 
\hat{Q} \, {Y_{J{\xi}} {\left(\hat Q\right)}  } \ - \ 
i\, \sqrt{ { {J} \over {2J+1}} } \, 
\hat{Q} \times {\vec Y^{\:{\xi}}_{JJ1} {\left(\hat Q\right)}  } ,
\label{id4}
\end{equation}
\begin{equation}
{\vec Y^{\:{\xi}}_{JJ-11} {\left(\hat Q\right)}  } \ = \ 
 \sqrt{ { {J} \over {2J+1}} } \, 
\hat{Q} \, {Y_{J{\xi}} {\left(\hat Q\right)}  } \ - \ 
i\, \sqrt{ { {J+1} \over {2J+1}} } \, 
\hat{Q} \times {\vec Y^{\:{\xi}}_{JJ1} {\left(\hat Q\right)}  } .
\label{id5}
\end{equation}

We use the identities (\ref{id4}) and (\ref{id5})
in the angular integrals of Eq.~(\ref{IMJLQ}) and get
\[
I^{\xi}_{JJ-1}(Q)=\int d{\hat Q\,'} {\vec I \left(\vec Q\,'\right)}
 {\vec Y^{\:{\xi}}_{JJ-11}{\left(\hat Q\,'\right)}} \ = 
\]
\begin{equation}\label{insid1}
\int d{\hat Q\,'} \left[
\sqrt{ { {J} \over {2J+1}} } \,
[ {\hat Q\,'}\cdot {\vec I \left(\vec Q\,'\right)} ]\, {Y_{J{\xi}} {\left(\hat Q\,'\right)} } \, + \,
i\, \sqrt{ { {J+1} \over {2J+1}} } \,
[{\hat Q\,'} \times {\vec I \left(\vec Q\,'\right)} ] \cdot {\vec Y^{\:{\xi}}_{JJ1} {\left(\hat Q\,'\right)}} \right] 
\end{equation}
\[
I^{\xi}_{JJ+1}(Q)=\int d {\hat Q\,'} {\vec I \left(\vec Q\,'\right)} 
{\vec Y^{\:{\xi}}_{JJ+11} {\left(\hat Q\,'\right)}} \ =
\]
\begin{equation}\label{insid2}
\int d {\hat Q\,'}  \left[
- \sqrt{ { {J+1} \over {2J+1}} } \,
[{\hat Q\,'} \cdot {\vec I \left(\vec Q\,'\right)} ]\, {Y_{J{\xi}} {\left(\hat Q\,'\right)} } \, + \,
i\, \sqrt{ { {J} \over {2J+1}} } \,
[{\hat Q\,'} \times {\vec I \left(\vec Q\,'\right)} ] \cdot {\vec Y^{\:{\xi}}_{JJ1} {\left(\hat Q\,'\right)}} \right] 
\end{equation}
Inserting Eqs.~(\ref{insid1}) and (\ref{insid2}) into Eq.~(\ref{el}) 
and making use of Eqs.~(\ref{id2}) and (\ref{id35}) leads to
\[
{\vec I^{\,el}_{J{\xi}} ({\vec Q} )} = {\hat Q}\, {Y^{\:*}_{J{\xi}}} {\left(\hat Q\right)} \, \int d {\hat Q\,'} [ {\hat Q\,'} \cdot {\vec I \left(\vec Q\,'\right)}  ]\, {Y_{J{\xi}}
{\left(\hat Q\,'\right)}} 
\]
\[
\ - \
 \, [ {\vec Y^{\:*{\xi}}_{JJ1} {\left(\hat Q\right)}} \times {\hat Q} ] \, 
\int d {\hat Q\,'} [ {\hat Q\,'} \times {\vec Y^{\:{\xi}}_{JJ1} {\left(\hat Q\,'\right)}} ] \, {\vec I \left(\vec Q\,'\right)}
\]
\begin{equation}
\equiv \ {\hat Q} \, {Y^{\:*}_{J{\xi}}} {\left(\hat Q\right)}  \, {4 \pi} \, 
{T^{el}_{J{\xi}}(Q;long)}
\ + \ i \, [ {\vec Y^{\:*{\xi}}_{JJ1} {\left(\hat Q\right)}  } \times {\hat Q} ] \, {4\pi} \, {T^{el}_{J{\xi}}(Q;transv)}   .
\end{equation}
In such a way we introduce the electric transverse
\begin{equation}
{T^{el}_{J{\xi}}(Q;transv)}  = {i\over4\pi Q}\int d {\hat Q\,'} \left[ {\vec Q\,'} \times
{\vec Y^{{\xi}}_{JJ1} {\left(\hat Q\,'\right)} } \right]\cdot {\vec I \left(\vec Q\,'\right)}  ,
\end{equation}
and the electric longitudinal multipoles
\begin{equation}
{T^{el}_{J{\xi}}(Q;long)}={1\over4\pi Q}\int d {\hat Q\,'} \left({\vec Q\,'}  \cdot
{\vec I \left(\vec Q\,'\right)} \right){Y_{J{\xi}}}
{\left(\hat Q\,'\right)} .
\end{equation}
The magnetic multipoles are defined accordingly
\begin{equation}
{T^{mag}_{J{\xi}}(Q)} ={1\over4\pi}\int d {\hat Q\,'} {\vec Y^{{\xi}}_{JJ1} {\left(\hat Q\,'\right)} }
 \cdot {\vec I \left(\vec Q\,'\right)}
 .
\end{equation}
Note that the magnetic multipoles are transverse since ${\vec Q}\cdot
{\vec Y^{{\xi}}_{JJ1} {\left(\hat Q\right)}}=0.$
The electric longitudinal multipoles ${T^{el}_{J{\xi}}(Q;long)}$
do not appear in our studies of the processes with real photons. 
With the help of the identity
\begin{eqnarray}
{\vec Q} \times {\vec Y^{{\xi}}_{JJ1} {\left(\hat Q\right)}}&=&
i \sqrt{ {J+1\over J}}\,{\vec Q}\,
{Y_{J{\xi}}}
{\left(\hat Q\right)} \nonumber
\\&+&i \sqrt{ {2J+1\over J}}\,Q\,{\vec Y^{{\xi}}_{J,J+1,1} {\left(\hat Q\right)} }
\end{eqnarray}
one can introduce a longitudinal term into the expression for the 
transverse electric multipoles.
The aim of this transformation is to get a contribution, which  can be 
associated
via the continuity equation
\begin{eqnarray}
{\vec Q} \cdot {\vec I \left(\vec Q\right)} &=&
\left<\vec P\,'\:\Psi^{(-)}_{f}
\left|\vphantom{\Psi^{(-)}_{f} }\left[H,\hat\rho(0)\right]\right|
\Psi_{{^{3}{He}}}\:\vec P\right> \nonumber
\\&=& \omega
\left<\vec P\,'\:\Psi^{(-)}_{f}
\left|\vphantom{\Psi^{(-)}_{f} }\hat\rho(0)\right|
\Psi_{{^{3}\!{He}}}\:\vec P\right>
=Q\rho\left({\vec Q} \right)
\end{eqnarray}
with the matrix elements of the charge density $\rho\left({\vec Q} \right)$.
So, the electric multipoles can be cast into the form
\begin{eqnarray}
 {T^{el}_{J{\xi}}(Q)} =-{1\over 4\pi\sqrt{J}}\int &d {\hat Q\,'} &\left[\sqrt{J+1}{Y_{J{\xi}}} 
{\left(\hat Q\right)} \rho
{\left(\vec Q\,'\right)}
\right.
\nonumber
\\&+&\left. \sqrt{2J+1} 
{\vec Y^{{\xi}}_{J,J+1,1} {\left(\hat Q\,'\right)}}\cdot 
{\vec I \left(\vec Q\,'\right)}\right] ,
\end{eqnarray} 
which is identical with Eq.~(\ref{eq:eq6}).

Finally, choosing the direction of the $z$--axis along 
the photon momentum we get for the
transverse components of the nuclear matrix element
\begin{equation}
\left[\vec I (Q\vec e_z) \right]_{\xi} \, = \,
-\sqrt{2\pi}\sum_{J>1}\sqrt{2J+1}\left[ {T^{el}_{J{\xi}}(Q)}+{\xi} {T^{mag}_{J{\xi}}(Q)}
\right], \ ({\xi}=\pm1),
\end{equation}
which coincides with Eq.~(\ref{eq:eq2}).

\begin{table}[hbt]
\begin{center}
\begin{tabular}{|r|r|r|}
\hline
 $E_p$ & $E_d$ & $Q$ \\
   MeV &   MeV & MeV/c \\
\hline
      5.0 &   10.0  &   8.8  \\
      8.8 &   17.5  &  11.4  \\
      9.9 &   19.8  &  12.1  \\
     14.8 &   29.6  &  15.4  \\
     22.5 &   45.0  &  20.5  \\
     47.5 &   95.0  &  37.2  \\
    100.0 &  199.9  &  72.3  \\
    150.0 &  299.9  & 105.7  \\
    200.0 &  399.8  & 139.1  \\
\hline
\end{tabular}
\end{center}
\caption[]
{Proton and deuteron lab energies and corresponding photon momenta
for the inverse reaction in the lab system for experiments 
analyzed in this work.}
\label{EpEdEgamma}
\end{table}

\end{document}